\documentclass[letterpaper]{article}

\usepackage{natbib,alifeconf}  
\usepackage{amssymb, amsmath}
\usepackage{url,hyperref,cleveref}
\usepackage{booktabs}
\usepackage{bm}
\usepackage{ dsfont }

\usepackage{graphicx}
\newcommand{\lowers}{{\scalebox{0.95}{$s$}}}

\usepackage{booktabs}
\usepackage{tabularx}
\usepackage{longtable}  

\title{Counting Short Trajectories in Elementary Cellular Automata using the Transfer Matrix Method}

\author{
Cédric Koller$^{1, \dagger}$ \and
Barbora Hudcová
$^{2}$
\mbox{}\\
$^1$Statistical Physics of Computation Laboratory, École Polytechnique Fédérale de Lausanne (EPFL), Switzerland \\
$^2$Chair of Statistical Field Theory, École Polytechnique Fédérale de Lausanne (EPFL), Switzerland \\
$^\dagger$Corresponding author: \href{mailto:cedric.koller@epfl.ch}{cedric.koller@epfl.ch}
} 

\begin{document}

\maketitle

\begin{abstract}
Elementary Cellular Automata (ECAs) exhibit diverse behaviours often categorized by Wolfram's qualitative classification. To provide a quantitative basis for understanding these behaviours, we investigate the global dynamics of such automata and we describe a method that allows us to compute the number of all configurations leading to short attractors in a limited number of time steps. This computation yields exact results in the thermodynamic limit (as the CA grid size grows to infinity), and is based on the Transfer Matrix Method (TMM) that we adapt for our purposes. Specifically, given two parameters $(p, c)$ we are able to compute the entropy of all initial configurations converging to an attractor of size $c$ after $p$ time-steps. By calculating such statistics for various ECA rules, we establish a quantitative connection between the entropy and the qualitative Wolfram classification scheme.
Class 1 rules rapidly converge to maximal entropy for stationary states ($c=1$) as $p$ increases. Class 2 rules also approach maximal entropy quickly for appropriate cycle lengths $c$, potentially requiring consideration of translations. 
Class 3 rules exhibit zero or low finite entropy that saturates after a short transient.
Class 4 rules show finite positive entropy, similar to some Class 3 rules.
This method provides a precise framework for quantifying trajectory statistics, although its exponential computational cost in $p+c$ restricts practical analysis to short trajectories.
\end{abstract}


\begin{small}Data/Code available at: 
\url{https://github.com/cedric-koller/counting_short_trajectories_ECA}.\end{small}

\section{Introduction}
Cellular automata (CAs) are discrete dynamical systems operating on a grid of cells, where each cell updates its state based on the states of its neighbors according to a local rule. Despite their structural simplicity, CAs can generate extraordinarily complex patterns and behaviors, leading to their use as models in physics \citep{chopard_cellular_1998}, biology \citep{ermentrout_cellular_1993}, and computer science (see e.g. \cite{bhattacharjee_survey_2020} for a survey). Elementary Cellular Automata (ECAs) are one of the simplest classes of CAs, operating on a 1D grid with binary states $\{0, 1\}$ and nearest-neighbor interactions \citep{wolfram_statistical_1983}. Despite the compactness of their local rules, elementary cellular automata can exhibit complex behaviors, and
one of them, rule 110, has been proven to be Turing complete \citep{cook_universality}.

\begin{figure}[t]
    \centering
    \begin{minipage}[b]{0.45\linewidth}
        \centering
        {\small Class 1\par}
        \vspace{0.3ex}
        \includegraphics[width=\linewidth]{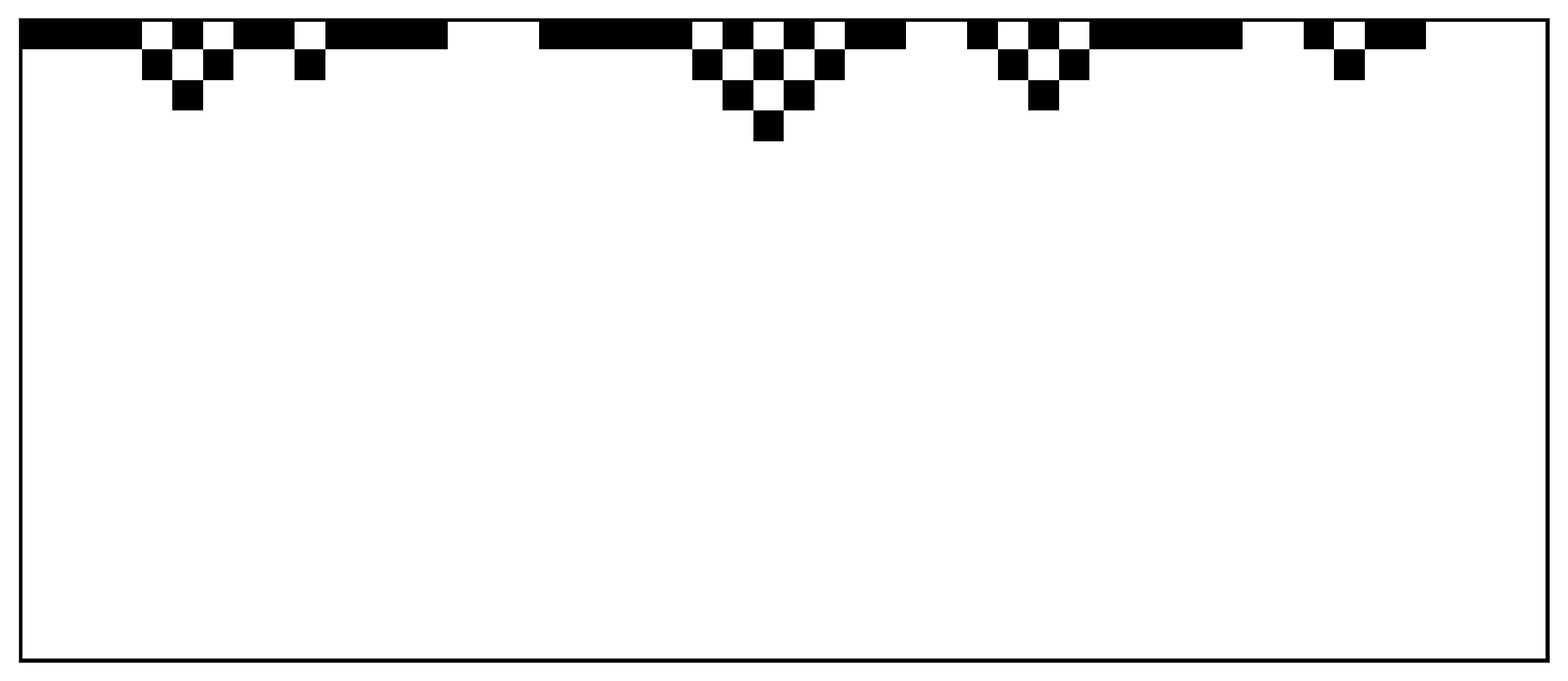}
    \end{minipage}
    \hspace{1em}
    \begin{minipage}[b]{0.45\linewidth}
        \centering
        {\small Class 2\par}
        \vspace{0.3ex}
        \includegraphics[width=\linewidth]{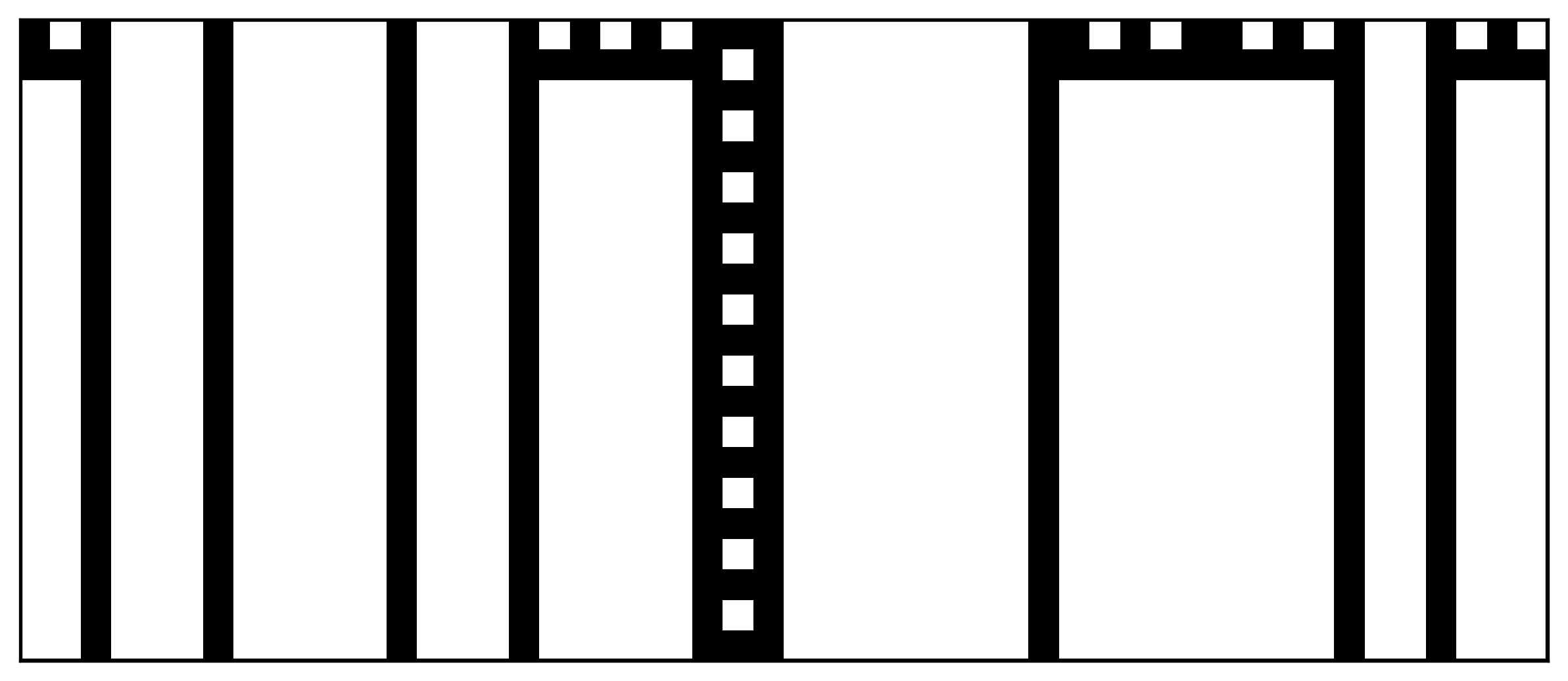}
    \end{minipage}

    \vspace{0.15em}

    \begin{minipage}[b]{0.45\linewidth}
        \centering
        {\small Class 3\par}
        \vspace{0.3ex}
        \includegraphics[width=\linewidth]{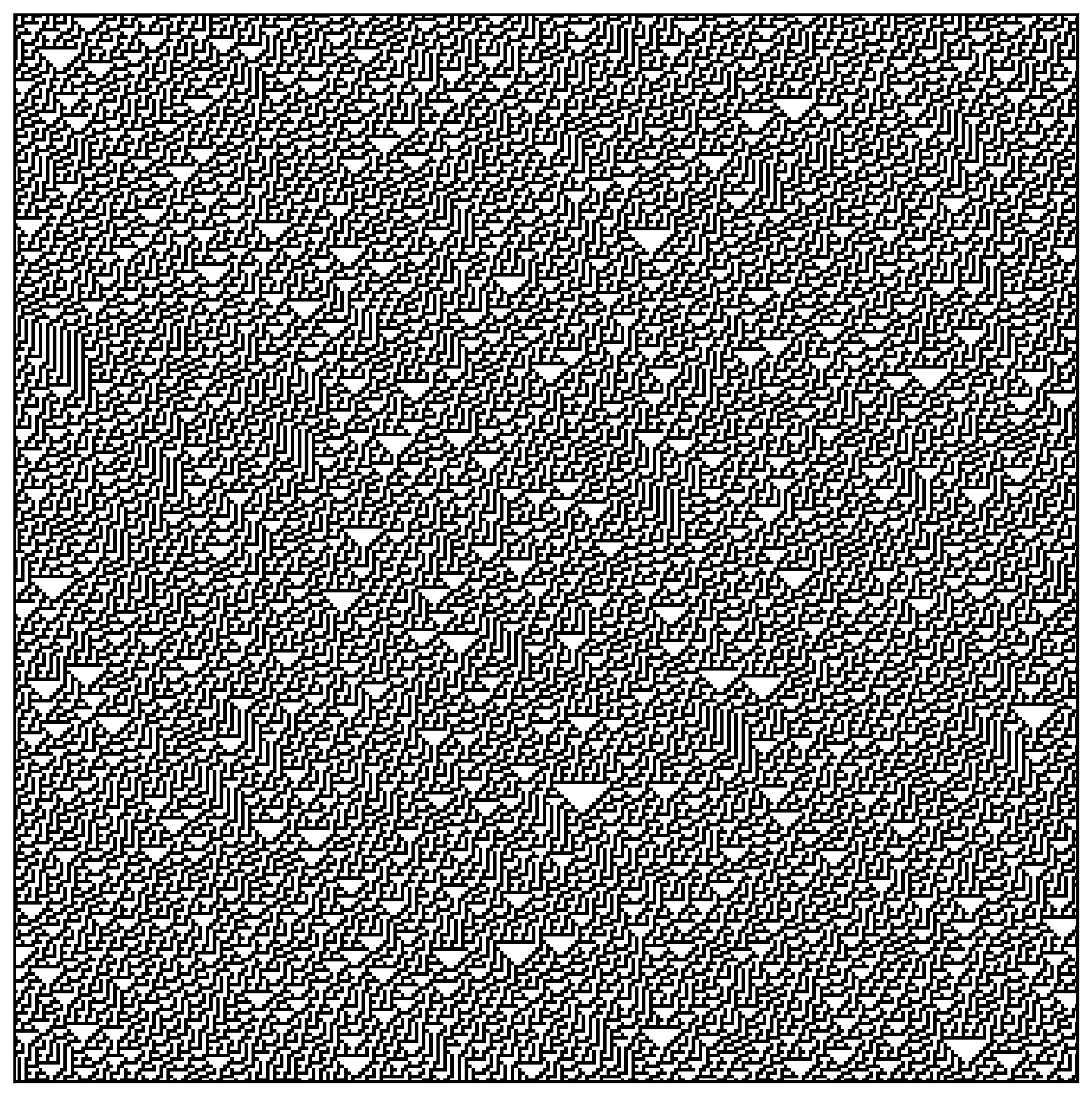}
    \end{minipage}
    \hspace{1em}
    \begin{minipage}[b]{0.45\linewidth}
        \centering
        {\small Class 4\par}
        \vspace{0.3ex}
        \includegraphics[width=\linewidth]{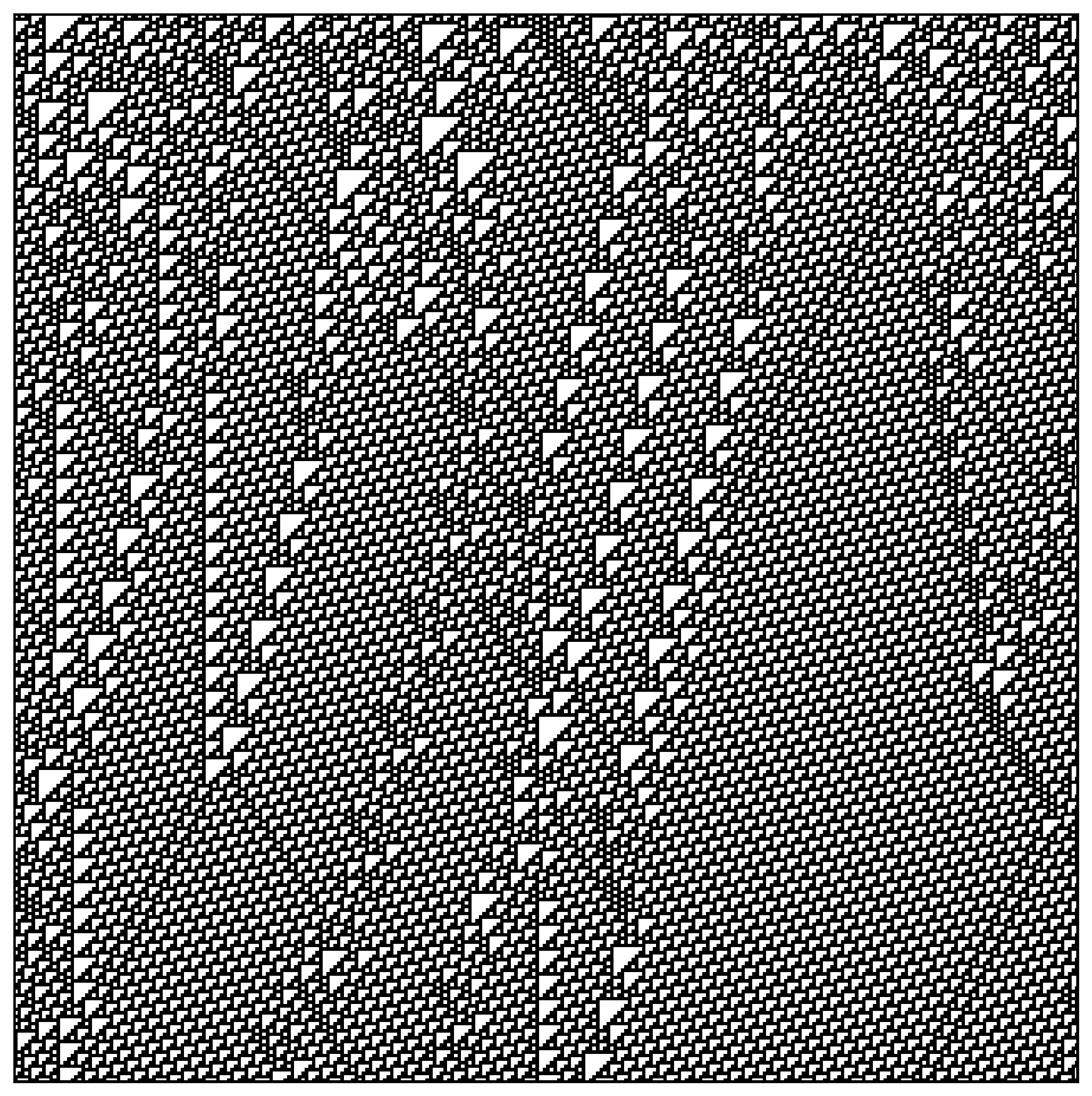}
    \end{minipage}

    \caption{Examples of space-time diagrams for Class 1 (rule 32), Class 2 (rule 108), Class 3 (rule 30), and Class 4 (rule 110) with random uniform initial configurations. On the horizontal axis: the configuration; on the vertical axis (top to bottom): evolution in time.}
    \label{fig:examples_wolfram_classes}
\end{figure}

\cite{wolfram_universality_1984} proposed a qualitative classification of ECA behavior based on their typical evolution from random initial conditions:
\begin{itemize}
  \item Class 1: Evolves to a homogeneous state (all 0s or all 1s).
  \item Class 2: Evolves to simple separated stable or periodic structures.
  \item Class 3: Evolves to chaotic, pseudo-random patterns.
  \item Class 4: Evolves to complex localized structures, sometimes long-lived.
\end{itemize}

The classification was made by observing \textit{space-time diagrams} which draw the configurations as horizontal rows of black and white squares (corresponding to $1$ and $0$ states) and the time evolution on the vertical axis, progressing downwards. Fig. \ref{fig:examples_wolfram_classes} shows examples of space-time diagrams for each class.

While insightful, the classification proposed by Wolfram is qualitative. Efforts have been made to establish more formal and quantitative measures of CA complexity. We briefly present some approaches below, and refer the reader to e.g. \cite{vispoel_progress_2022} for a more complete overview.

\cite{culik1988undecidability} proposed a formal definition of the Wolfram classes, and showed that such a classification is  undecidable. Indeed, the properties concerning the asymptotic dynamics of CAs are
in general undecidable \citep{kari1994reversibility,kari_rices_1994,  blanchard_topological_1997, delacourt_2021}. However, this does not mean that nothing can be said in specific scenarios or in average cases. Also note that alternative classifications have been proposed, see e.g. \cite{martinez_note} for an overview.

\cite{langton1990computation} introduced the Langton parameter, given by the fraction of neighborhood configurations that do not map to a quiescent state. This parameter roughly partitions ECAs according to Wolfram's classification. In a related effort, \cite{Wuensche_global_2001} proposed the so-called $Z$ parameter, defined as the probability that the next unknown cell in a partial preimage (a partially specified configuration in the previous time-step) is not determined. Class 4 typically appears at a fixed value of $Z$ (0.75 for ECAs). However, these parameters are very coarse since they relate to the property of the CA's local rule, disregarding its iterative application.

\cite{zenil_compression-based_2010} proposed a classification based on the compressibility of space-time diagrams, where complex rules are assumed to generate less compressible patterns. However, this method's results are sensitive to simulation parameters like the initial configuration and run time. This can lead to inconsistent classifications, as noted by the author and demonstrated in \cite{Hudcova_classification_2022}.

\cite{marr_topology_2005, marr_cellular_2012} related Wolfram's classes to the Shannon and the word entropy (the entropy of blocks of various sizes). However, the boundaries between the classes are determined \textit{a posteriori}, are not clear-cut, and vary depending on the rule. A more general discussion of the relationship between information theory and complexity can be found in \cite{prokopenko2009information}. \cite{santamaria2017package} proposes a package for measuring complexity based on the Shannon entropy, and \cite{lopez2023temporal} applies a similar approach to random Boolean networks.

A more recent effort is based on transient dynamics before the system settles into an attractor (a fixed point or a cycle). The transient length is the number of time steps until a finite system enters a cycle or stationary state. \cite{Hudcova_classification_2022} proposed classifying discrete dynamical systems based on how this transient length scales. Intuitively, simple systems (like Classes 1 \& 2) are expected to have short transients, while chaotic (Class 3) or complex (Class 4) systems exhibit longer transients before repeating.

The approaches presented above, except for the Langton and $Z$ parameters, rely on statistics computed from multiple simulations of the dynamics. We aim to obtain a metric that does not depend on the choice of the initial condition but describes the average behavior of typical initial conditions. Recently, \cite{behrens_backtracking_2023, behrens_dynamical_2023} investigated the average behavior of short trajectories in CAs on sparse graphs using the cavity method. They focused on counting configurations that evolve through a specific transient phase of length $p$ followed by a cycle of length $c$. This provides a detailed characterization of the paths that lead to the attractors. 

The cavity method can be used in 1D systems such as ECAs, but a more direct and conceptually straightforward path, which yields the same results, is the transfer matrix method (TMM) \citep{kramers_statistics_1941_1, kramers_statistics_1941_2, Baxter_Exactly_1984}. The transfer matrix method is a standard tool in statistical mechanics, particularly well-suited for the analysis of 1D systems. In this work, we adapt the method to compute the entropy density associated with such $(p, c)$-trajectories in ECAs. The TMM allows for exact calculations in the thermodynamic limit ($N\rightarrow \infty$, where $N$ is the system size). We compute the entropy density $s$ which measures the exponential growth rate of the number of initial configurations leading to a $(p, c)$-trajectory for a given initial density of active cells. We analyze how $s$ behaves as a function of $p$ and $c$ for representative rules from each Wolfram class, providing a quantitative perspective on their dynamic properties.

\section{Background and notation}
\paragraph{Elementary Cellular Automata}
We consider elementary cellular automata on a cyclic grid with $N\in \mathbb{N}$ sites. Each site $i$ is in a state $s^i_t\in\{0,1\}$ at each discrete time step $t$. The system evolves according to a deterministic local rule $f:\{0,1\}^3\rightarrow \{0,1\}$, such that the state of site $i$ at time $t+1$ is given by
\begin{equation}
    s^i_{t+1}=f(s^{i-1}_t, s^i_t, s^{i+1}_t) \quad\forall\,i=1,\hdots, N.
\end{equation}
Since the grid is cyclic, we have periodic boundary conditions $s^0_t=s^N_T$ and $s^{N+1}_t=s^1_t$. Let $\bm s_t=(s^1_t, \hdots, s^N_t)\in\{0,1\}^N$ denote the \textit{configuration} at time $t$. The global evolution map $F:\{0,1\}^N \rightarrow \{0,1\}^N$
is given by the synchronous update
\begin{equation}
F(\bm s_t)=(f(s^N_t, s^1_t, s^2_t),\hdots, f(s^{N-1}_t, s^N_t, s^1_t))=\bm s_{t+1}.
\end{equation}
To each local update function is associated the \textit{Wolfram number} $k$ defined as
\begin{equation}
k=2^0f(0,0,0)+2^1f(0,0,1)+\hdots +2^7f(1,1,1).
\end{equation}
For example, the identity update function that keeps every configuration unchanged has the Wolfram number $2^0\cdot 0+2^1\cdot 0+2^2\cdot 1+2^3\cdot 1+2^4\cdot 0+2^5\cdot 0+2^6\cdot 1+2^7\cdot 1=204$.
We refer to an ECA with the update function associated with the Wolfram number $k$ as ``rule $k$". There are $256$ different ECA rules. Removing the ``left $\leftrightarrow$ right" and ``$0\leftrightarrow 1$" symmetries, $88$ non-equivalent rules remain \citep{li1990structure, SCHALLER2025100298} and will be studied in this work.


\paragraph{Counting trajectories} We denote a \textit{sequence of configurations} as $\underline{\bm s}=(\bm s_1, \hdots, \bm s_T)$ . We write $\underline{s}^i=(s^i_1, \hdots, s^i_T)$ the \textit{sequence of site} $i$. We denote as $\underline{\bm S}$ the ensemble of all possible sequences of configurations of total duration $T\in \mathbb{N}$, and as $\underline{S}=\{0,1\}^T$ the ensemble of all sequences for a single site. A \textit{trajectory} $\underline{\bm s}$ is a sequence of configurations that respects the evolution map $F$, i.e. $F(\bm s_t)=\bm s_{t+1}\, \forall t=1, \hdots, T-1$. We will call the total duration $T$ the \textit{total length} (or just \textit{length}) of a trajectory. A configuration $\bm s$ is said to be \textit{stationary} if $F(\bm s)=\bm s$ and \textit{homogeneous} if $s^1 = s^2 = \hdots = s^N$.

We study trajectories that have a transient length $p\in\mathbb{N}$ and a cycle length $c\in\mathbb{N}$, i.e. trajectories that have a total length $T=p+c$ such that $F(\bm s_{p+c})=\bm s_{p+1}$. Our goal is, given $p$, $c$ and a fixed ECA rule, to compute the number of initial configurations that, after $p$ transient steps, enter a cycle of length $c$. For this, we will leverage some standard tools of statistical physics. We refer the interested reader to \cite{yeomans_statistical_1992} for an introduction to statistical physics, in particular chapter 5 for the transfer matrix method. Concretely, for a fixed system size $N$, we introduce a probability distribution over all sequences of configurations of length $p+c$ such that valid trajectories have a non-zero probability, and other sequences of configurations have a probability $0$. Quantities of interest related to this distribution, such as the logarithm of the number of valid initial configurations, can then be derived using the transfer matrix method.

\begin{figure}
    \centering
    \includegraphics[width=0.45\textwidth]{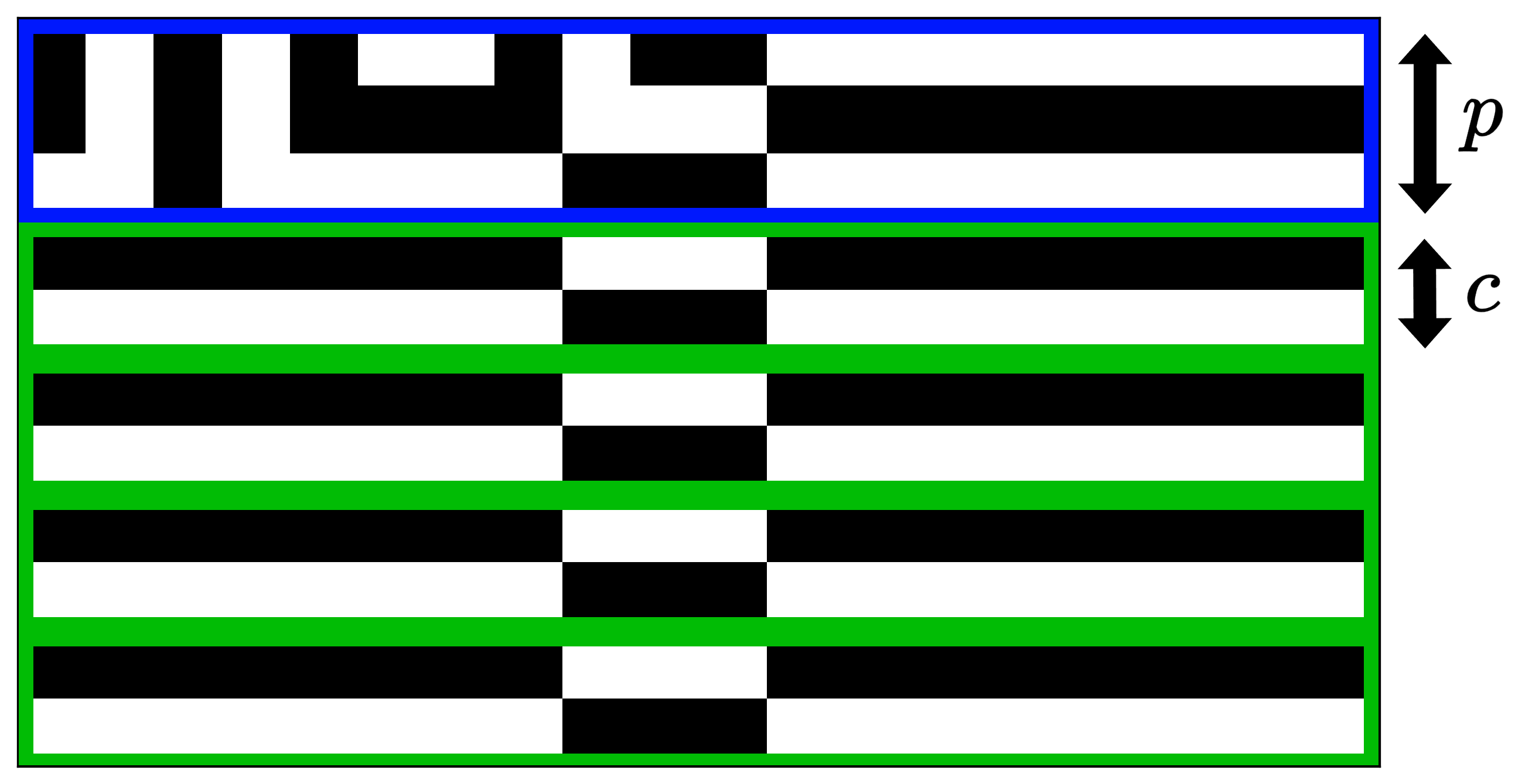} 
    \caption{Example of a trajectory of rule 23 with a transient of length $p=3$ (blue) and a cycle of length $c=2$ (green) repeated 4 times.}
    \label{fig:example_p_c}
\end{figure}

The probability distribution reads
\begin{small}
\begin{equation}\label{eq:probability measure}
P(\underline{\bm s})=\frac{e^{\mu \sum_{i=1}^N s_1^i}}{Z} \mathds{1}\left[F(\bm s_{p+c})=\bm s_{p+1}\right] \prod_{t=1}^{p+c-1}\mathds{1}\left[F(\bm s_t)=\bm s_{t+1}\right],
\end{equation}
\end{small} 
where $\mathds{1}$ is the indicator function that is $1$ if the statement in brackets is true and $0$ otherwise. $Z$ is the normalization constant or \textit{partition function} of the probability distribution:
\begin{small}
\begin{equation}\label{eq:partition function}
    Z=\sum_{\underline{\bm s} \in \underline S} e^{\mu \sum_{i=1}^N s_1^i}\mathds{1}\left[F(\bm s_{p+c})=\bm s_{p+1}\right] \prod_{t=1}^{p+c-1}\mathds{1}\left[F(\bm s_t)=\bm s_{t+1}\right].
\end{equation}
\end{small}
The first indicator function ensures that the trajectory indeed ends in a cycle of length $c$, and the product of indicator functions ensures that the sequence of configurations respects the global update rule $F$, i.e. that it is a proper trajectory. We use the nomenclature of \cite{behrens_backtracking_2023} and call these trajectories \textit{backtracking attractors}. Figure \ref{fig:example_p_c} shows an example of a backtracking attractor with $p=3$ and $c=2$. Note that cycles are allowed within the transient phase, as our focus is on quantifying initial configurations that ultimately transition out of this phase into a cycle.

The parameter $\mu \in \mathbb{R}$ is added to tilt the measure towards specific initial densities. If $\mu=0$, $P(\underline{\bm s})$ defined in \eqref{eq:probability measure} becomes the uniform probability distribution over trajectories of length $p+c$ following the global rule $F$ and ending in a cycle of length $c$. In that case, $Z$ is the number of trajectories that obey these constraints.

The transfer matrix method will allow us to compute the \textit{free entropy density}
\begin{equation}\label{eq:free entropy}
    \phi=\frac{\log Z}{N}
\end{equation}
in the limit $N\rightarrow \infty$. The free entropy density allows to compute other quantities of interest. For instance, the expected initial density of alive cells
\begin{equation}
    \rho=\mathbb{E}_{\underline{\bm s} \sim P}\left[\frac{\sum_{i=1}^N s_i^1}{N}\right]
\end{equation} is obtained as
\begin{equation}\label{eq:density}
    \rho=\frac{\partial \phi}{\partial \mu}.
\end{equation}
Thus, $\mu$ is the \textit{Lagrange multiplier} associated to $\rho$. Factorizable quantities (quantities that can be written as $\sum_{i=1}^N g(\underline s^{i-1}, \underline s^i)$ for $g$ an arbitrary function) can be obtained using other Lagrange multipliers. Stochasticity can also be introduced with a temperature-like Lagrange multiplier. However, we focus in this work only on the initial density $\rho$ and deterministic evolution. Note that it is expected that these quantities are self-averaging, i.e. concentrate around their mean.

Let $\mathcal{N}(\rho)$ be the number of backtracking attractors with initial density $\rho$. We define the \textit{entropy density} $s$ as the logarithm of the number of attractors divided by $N$:
\begin{equation}
    s=\frac{\log( \mathcal{N}(\rho))}{N}.
\end{equation}
Using the saddle-point method in the limit $N\rightarrow \infty$, the entropy density can also be obtained from the free entropy density as
\begin{equation}
    s=\phi-\mu\rho.
\end{equation}


Readers unfamiliar with statistical physics should interpret the entropy as a proxy for the number of initial conditions leading to ($p,c$)-attractors. This is to be distinguished from the Shannon entropy, measuring the randomness of configurations. Importantly, calculating the total entropy for multiple attractors requires summing the number of initial states, not the individual entropy values.

\section{Transfer Matrix for ECA}
We first note that \cite{lemoy_transfer_2014} studied the stationary configurations of noisy 1-dimensional majority cellular automata using the transfer matrix method. We extend this formalism to the dynamic case for any ECA rule. Dynamical transfer matrices were already used, e.g. in \cite{coolen_transfer_2012} to study random field and bond Ising models. However, the study of backtracking attractors using the transfer matrix is novel to our knowledge. The transfer matrix method offers asymptotically exact results under certain technical conditions that we will not detail here for brevity \citep{yeomans_statistical_1992, horn2012matrix}. We will in particular assume that the largest eigenvalue of the transfer matrix is non-degenerate and real.

The partition function \eqref{eq:partition function} can be written in a factorized form as
\begin{equation}\label{eq:partition function factorized}
\begin{aligned}
Z=\sum_{\underline{\bm s}\in \underline{S}} &
\prod_{i=1}^N e^{ \mu s_1^i}
\mathds{1}\left[f(s_{p+c}^{i-1}, s_{p+c}^{i}, s_{p+c}^{i+1})=\bm s_{p+1}^i\right]
\\
&\times
\prod_{t=1}^{p+c-1}\mathds{1}\left[f(s_t^{i-1}, s_t^{i}, s_t^{i+1})=\bm s_{t+1}^i\right].
\end{aligned}
\end{equation}

Our aim is to write the partition function in terms of the \textit{transfer matrix} $\mathcal{T}\in \mathbb{R}^{4^{p+c}\times4^{p+c}}$.
The transfer matrix is defined with its components:
\begin{equation}
\begin{aligned}
\mathcal{T}_{(\underline{\lowers}^{\alpha}, \underline{\lowers}^{\beta}), (\underline{\sigma}^\alpha, \underline{\sigma}^{\beta})}
    = 
e^{ \mu s_1^i}
\mathds{1}\left[f(s_{p+c}^{\alpha}, s_{p+c}^{\beta}, \sigma_{p+c}^{\beta})= s_{p+1}^{\beta}\right]
\\
\times
\mathds{1}\left[\underline{s}^\beta=\underline{\sigma}^\alpha\right] \times
\prod_{t=1}^{p+c-1}\mathds{1}\left[f(s_t^{\alpha}, s_t^{\beta}, \sigma_t^{\beta})=s_{t+1}^\beta\right].
\end{aligned}
\end{equation}
Each row index corresponds to a tuple of sequences of site $(\underline{s}^\alpha, \underline{s}^\beta)$ that correspond to an (arbitrary) site $i-1$ and $i$ respectively. Similarly, each column is indexed by $(\underline{\sigma}^\alpha, \underline{\sigma}^\beta)$ and represents the sequences at sites $i$ and $i+1$. The case $\underline{s}^\beta \neq \underline{\sigma}^\alpha$ is enforced to be $0$ by an indicator function.
We use the following arbitrary convention for the mapping between the tuples and the indexes:
\begin{scriptsize}
\begin{equation}
\begin{aligned}
    &\big( (0, \hdots,0, 0), (0,\hdots,0, 0) \big) \rightarrow 1
    , \big( (0, \hdots,0, 0), (0,\hdots,0, 1) \big) \rightarrow 2,
    \\
    &\big( (0, \hdots,0, 1), (0,\hdots,0, 0) \big) \rightarrow 3, \big( (0, \hdots,0, 1), (0,\hdots,0, 1) \big) \rightarrow 4,
    \\
    &\big( (0, \hdots,0, 0), (0,\hdots,1, 0) \big) \rightarrow 5,
    \hdots,
    \big( (1, \hdots,1, 1), (1,\hdots,1, 1) \big) \rightarrow 4^{p+c}.
\end{aligned}
\end{equation}
\end{scriptsize}
As an example, take the case $p=0$, $c=1$. Then, the transfer matrix reads
\begin{equation}
\begin{tiny}
\mathcal{T}
=
\begingroup 
\setlength\arraycolsep{1pt}
\begin{pmatrix}
    \mathds{1}\left[ f(0,0,0)=0 \right]
    &
    \mathds{1}\left[ f(0,0,1)=0 \right]
    &
    0
    &
    0
    \\
    0
    &
    0
    &
    e^{\mu}    \mathds{1}\left[ f(0,1,0)=1 \right]
    &
    e^{\mu} \mathds{1}\left[ f(0,1,1)=1 \right]
    \\
    \mathds{1}\left[ f(1,0,0)=0 \right]
    &
    \mathds{1}\left[ f(1,0,1)=0 \right]
    &
    0
    &
    0
    \\
    0
    &
    0
    &
    e^{\mu} \mathds{1}\left[ f(1,1,0)=1 \right]
    &
    e^{\mu} \mathds{1}\left[ f(1,1,1)=1 \right]
\end{pmatrix}
\endgroup.
\end{tiny}
\end{equation}

The partition function \eqref{eq:partition function factorized} can be written as
\begin{equation}
    Z=\sum_{\underline{\bm s}\in \underline{S}}
    \prod_{i=1}^N \mathcal{T}_{(\underline{\lowers}^{i-1}, \underline{\lowers}^i),(\underline{\lowers}^i, \underline{\lowers}^{i+1})}.
\end{equation}
We have that $\text{Tr}\left( \mathcal{T}^N\right)=Z$.
Indeed,
\begin{equation}
\begin{scriptsize}
\begin{aligned}
\text{Tr}\left(\mathcal{T}^N
\right)
&=
\sum_{(\underline{s}^1, \underline{\sigma}^2)\in \underline{S}^2}\left( \sum_{(\underline{s}^2, \underline{\sigma}^3)\in\underline{S}^2}\hdots \sum_{(\underline{s}^N, \underline{\sigma}^1)\in \underline{S}^2}
\prod_{i=2}^{N+1}
\mathcal{T}_{(\underline{s}^{i-1}, \underline{\sigma}^{i}), (\underline{s}^{i}, \underline{\sigma}^{i+1})}
\right)
\\
&=
\sum_{\underline{\bm s}\in \underline{\bm S}} \sum_{\underline{\bm \sigma}\in \underline{\bm S}}\prod_{i=1}^N \mathcal{T}_{(\underline{s}^{i-1}, \underline{\sigma}^{i}), (\underline{s}^i, \underline{\sigma}^{i+1})}
=Z,
\end{aligned}
\end{scriptsize}
\end{equation}
where the simplification in the last equality comes from the term $\mathds{1}\left[\underline{s}^i=\underline{\sigma}^i\right]$ included in $\mathcal{T}$, which implies that the product over $i$ is zero unless $\underline{\bm s}=\underline{\bm \sigma}$. We also used the periodic boundary conditions $\underline{s}^{N+2}=\underline{s}^2$.

For any square matrix $\mathcal{T}$, we have the relation $\text{Tr}\left(\mathcal{T}^N\right)=\sum_k \lambda_k^N$ where $\lambda_k$ is the $k$-th eigenvalue of $\mathcal{T}$ and the sum is over all eigenvalues. We can thus write the normalization as a function of the eigenvalues of the transfer matrix: $Z=\sum_{k}\lambda_k^N$.
The free entropy density \eqref{eq:free entropy} then reads
\begin{equation}
    \phi=\frac{\log\left(\sum_k \lambda_k^N\right)}{N}.
\end{equation}
In the limit $N\rightarrow \infty$, the free entropy density is dominated by the largest eigenvalue, so that we have
\begin{equation}
    \phi=\log\left(\lambda_{\text{max}}\right),
\end{equation}
where $\lambda_{\text{max}}$ is the largest eigenvalue of $\mathcal{T}$.
Thus, the free entropy density can be computed from the maximum eigenvalue of a $4^T\times 4^T$ matrix instead of the naive summation over the $(2^T)^N$ possible sequences of configurations.

The derivative of eq. \eqref{eq:density} can be computed using the Hellmann-Feynman theorem as
\begin{equation}
\frac{\partial \phi}{\partial \mu}=\frac{1}{\lambda_{\text{max}}}\frac{\partial \lambda_{\text{max}}}{\partial \mu}
=
\frac{\bm{v_L}^T \frac{\partial \mathcal{T}}{\partial \mu} \bm{v_R}}{\lambda_{\text{max}}\bm{v_L}^T\bm{v_R}},
\end{equation}
where $\bm{v_R}$ and $\bm{v_L}$ are the right and left eigenvectors associated with the largest eigenvalue $\lambda_{\text{max}}$, and the derivative $\frac{\partial \mathcal{T}}{\partial \mu}$ is taken element by element. In our case, the derivative is simply obtained by multiplying each element of $\mathcal{T}$ by $s^i_1$.

\vspace{-10pt} 

\paragraph{Technical implementation}
The transfer matrices are sparse, as the deterministic evolution forbids many sequences of configurations. We used the implicitly restarted Arnoldi algorithm \citep{sorensen_implicit_1992, sorensen_implicitly_1997} implemented in the ARPACK package \citep{ARPACK} via the SciPy library \citep{2020SciPy-NMeth} to compute the largest eigenvalue and associated right and left eigenvectors. To compute the entropy corresponding to a prescribed initial density $\rho$, we perform a bisection search on the parameter $\mu$ until the resulting density is within a specified tolerance $\epsilon=10^{-3}$.

\section{Results and Discussion}

\cite{wolfram_universality_1984} proposed a classification of ECA rules depending on their apparent complexity. However, as previously discussed, this classification is qualitative and may vary depending on the source. For example, rules 60, 90, 105, and 150 are categorized as Class 3 by the Wolfram Alpha platform\footnote{The platform is accessible at \url{https://www.wolframalpha.com/}} (as compiled in \cite{alfaro_classification_2024}), whereas they are classified as Class 2 in \cite{Hudcova_classification_2022}. Similarly, rules 72, 104, 200, and 232 are assigned to Class 1 in \cite{Hudcova_classification_2022}, despite typically evolving toward stationary configurations that are not homogeneous. In this work, we adopt this last convention, even though our approach can distinguish homogeneous and non-homogeneous stationary states by probing the density. For the remaining rules, we indicate both classifications in Table \ref{tab:rule-summary}.

\vspace{-15pt} 

\paragraph{Class 1}

\begin{figure}
    \centering
    \includegraphics[width=0.4\textwidth]{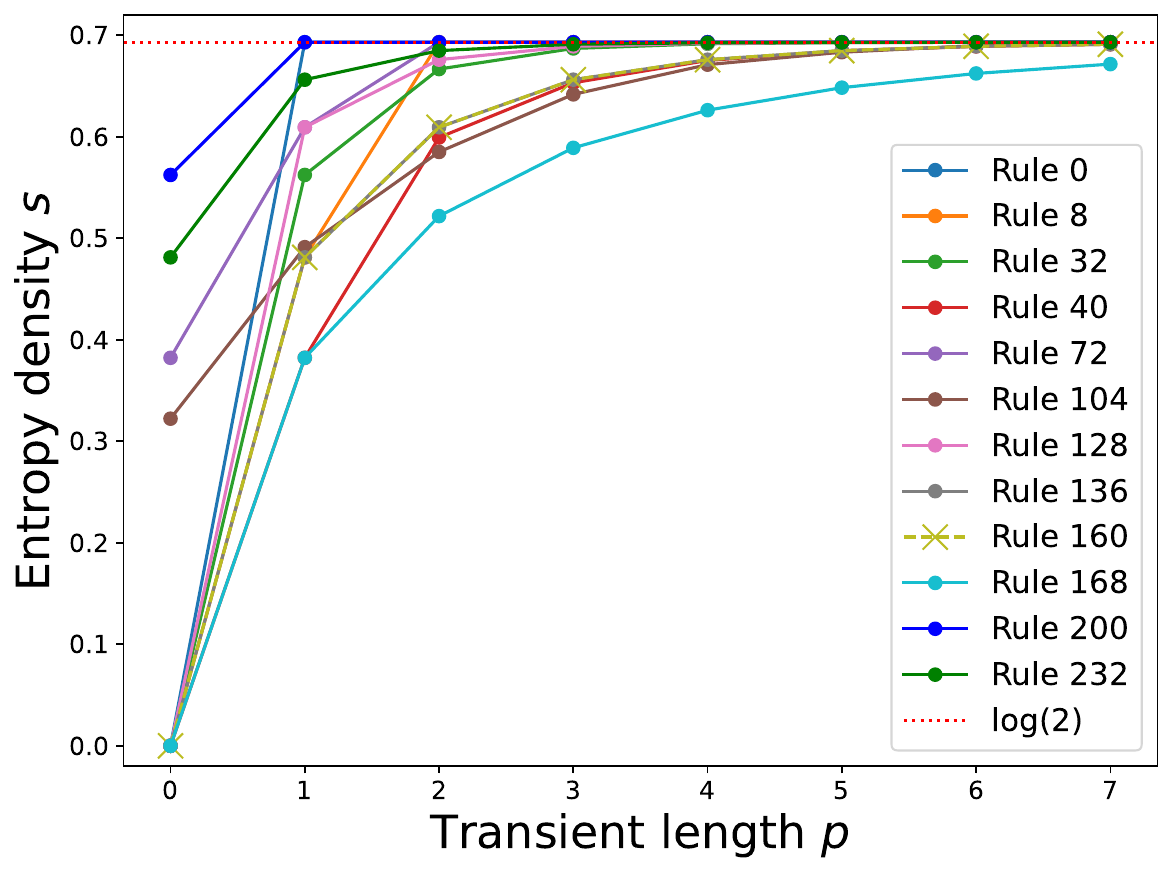} 
    \caption{Entropy density $s$ as a function of the transient length for trajectories ending in stationary configuration ($c=1$) for all Class 1 rules. The entropies are strictly increasing and quickly approach $\log(2)$ with increasing $p$. Rule 136 presents the same entropy profile as rule 160.}
    \label{fig:class_1}
\end{figure}

Rules from Class 1 quickly evolve to a stationary configuration (a configuration that does not evolve over time). This means that the entropy density of backtracking attractors with $c=1$ should quickly increase as $p$ increases. Figure \ref{fig:class_1} shows the entropies obtained from the TMM for increasing $p$'s. The entropy is indeed quickly increasing, and plateaus at $\log(2)$. This is the maximal possible entropy density, as there are $2^N$ possible initial configurations, and the dynamic is deterministic so that there are $2^N$ possible trajectories. This quantitatively confirms the intuition that these systems quickly erase initial complexity, evolving towards a simple stationary state for almost all initial configurations.

\begin{figure}
    \centering
    \includegraphics[width=0.388\textwidth]{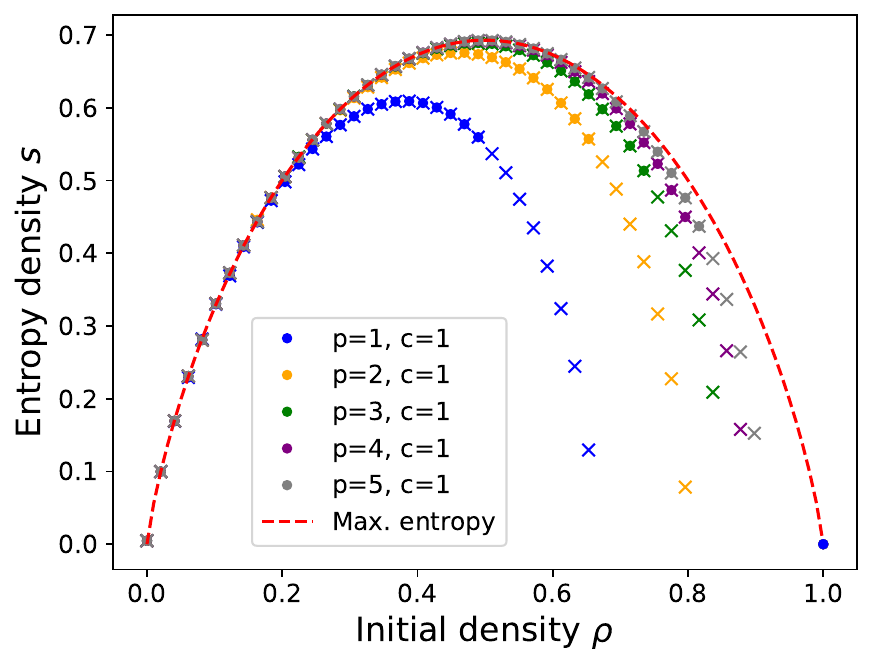} 
    \caption{Entropy density $s$ as function of the density $\rho$ for rule $128$ and various transient lengths $p$. Crosses indicate the entropy obtained for trajectory ending in the all $0$ configuration, while dots have no constraint on the reached stationary configuration.}
    \label{fig:entropy_density}
\end{figure}

Fig. \ref{fig:entropy_density} depicts the entropy density $s$ as a function of the initial density $\rho$, derived by varying the Lagrange multiplier $\mu$. This figure brings insights into the system's dynamics. For instance, the proximity of $s$ to the maximum theoretical entropy for lower initial densities indicates that initial configurations with low density of alive cells typically quickly evolve to a stationary configuration. This example also demonstrates behavior analogous to a first order phase transition. As $\mu$ increases for the case $p=1$, $c=1$, the obtained initial density $\rho$ abruptly transitions from $\rho=0.5$ to $\rho=1$ (see blue dots in Fig. \ref{fig:entropy_density}). This phenomenon arises from a shift in the dominant contribution to the partition function: for $\mu$ exceeding a critical value, the trajectory consisting only of $1$'s (which necessitates an initial density $\rho=1$) becomes thermodynamically dominant. However, a simple analysis of rule 128 confirms that initial configurations with initial densities up to $\rho=2/3$ exist and evolve in $p=1$ step to an all $0$ stationary configuration. These initial configurations are subdominant and do not contribute to the entropy. To access these subdominant contributions, one can restrict the probability distribution \eqref{eq:probability measure} to trajectories ending in the all $0$ configuration. This constraint effectively removes the dominant contribution from the all $1$ initial state. Under this modified ensemble, the full spectrum of initial densities up to $\rho=2/3$ is recovered, as indicated by the crosses in Fig. \ref{fig:entropy_density}.


\vspace{-15pt} 

\paragraph{Class 2}

\begin{figure}[!b]
    \centering
    \includegraphics[width=0.22\textwidth]{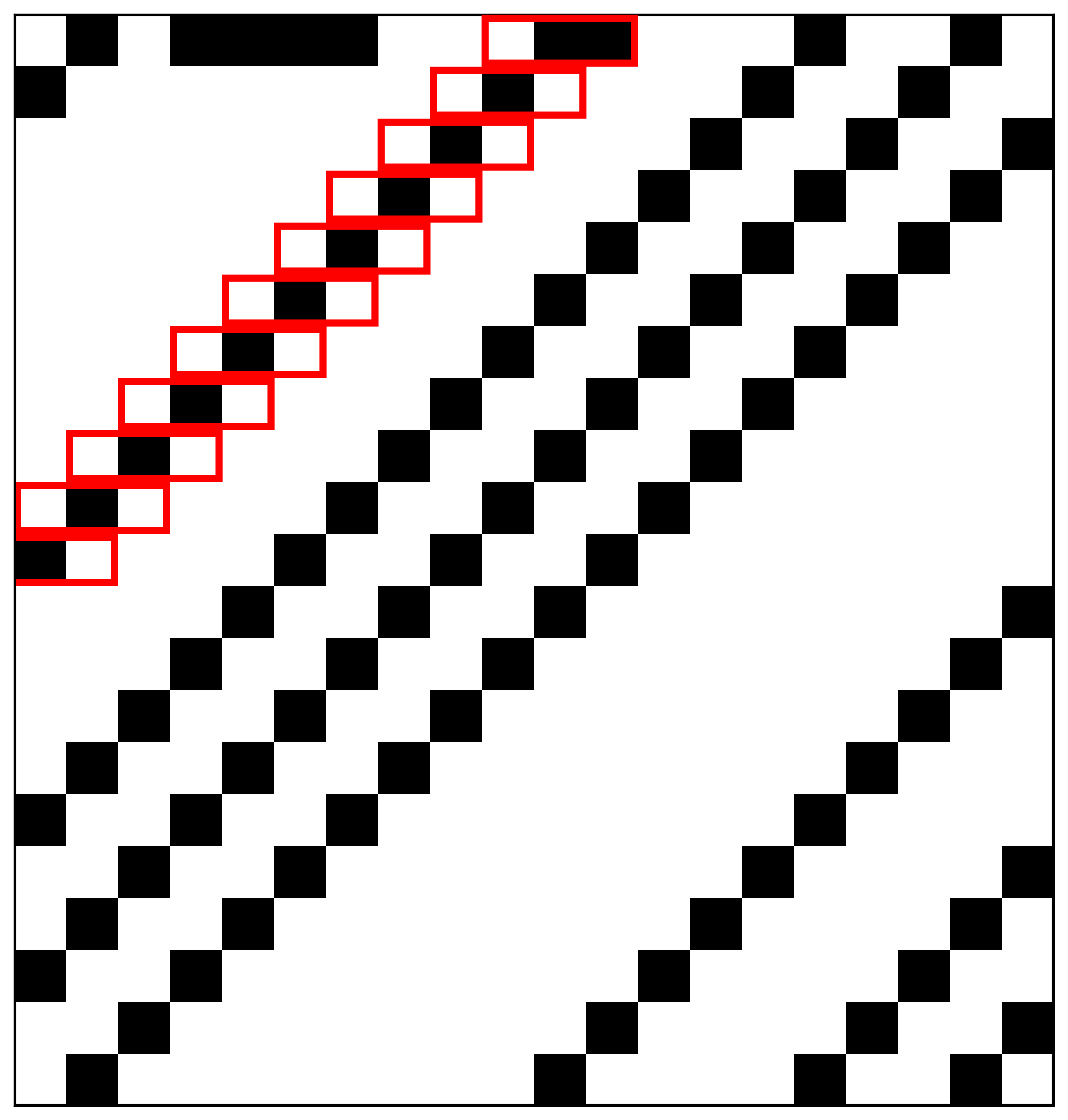} 
    \quad
    \includegraphics[width=0.22\textwidth]{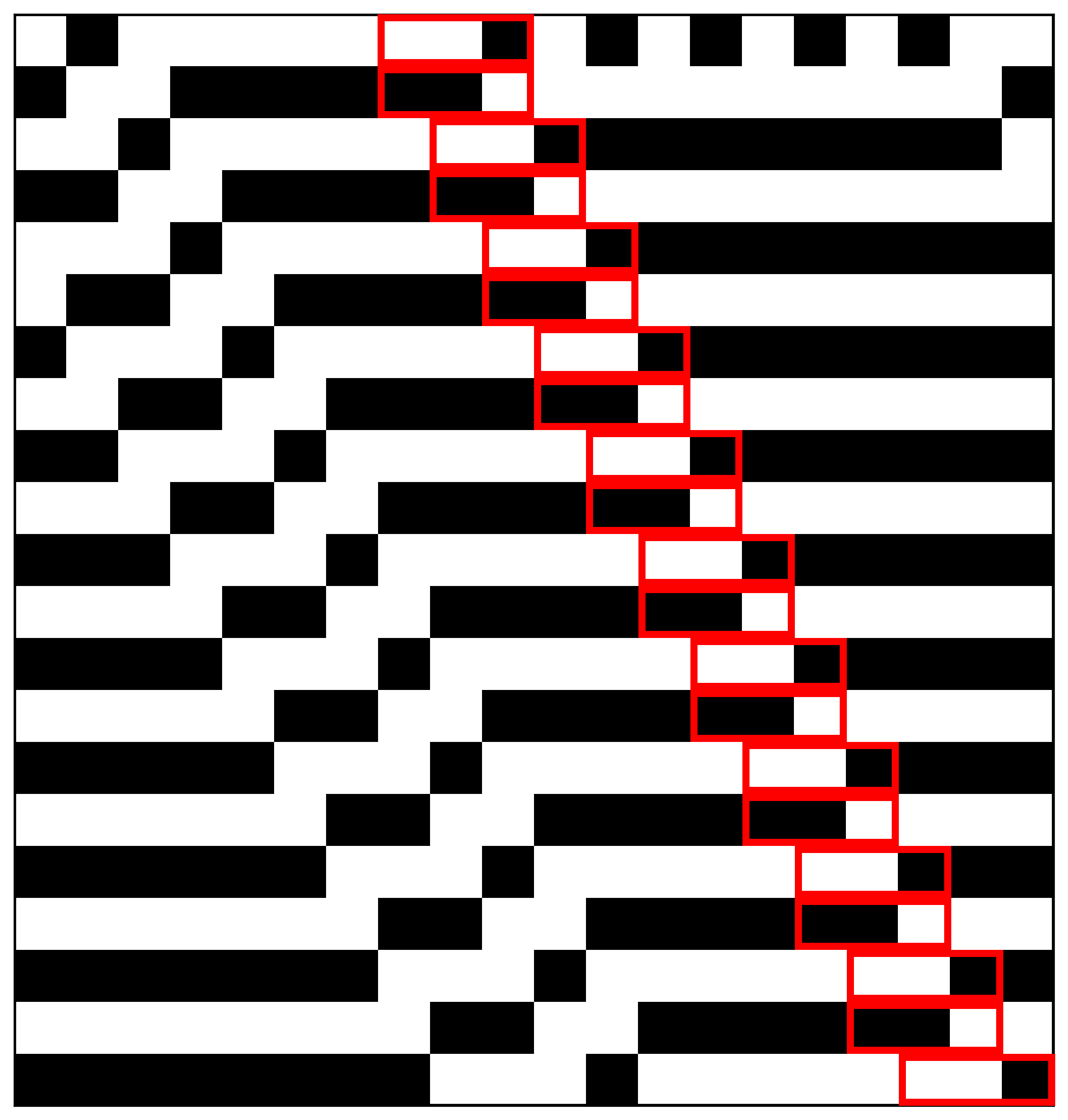} 
    \caption{Left: Example trajectory of rule 2. This rule presents a simple repeated but translated pattern. To take this into account, we use the \textit{left-translated} neighborhoods (drawn in red for a fixed cell at an arbitrary position $i$ at time $t=1$). Right: Similarly but for rule 3 and \textit{even-time right translated} neighborhoods.}
    \label{fig:shifting}
\end{figure}

Class 2 rules lead to a set of separated and simple stable or periodic structures. We note that the periodicity in time might be accompanied by spatial translations (see Fig. \ref{fig:shifting} for examples). Counting trajectories that end in cycles would give a low entropy in this case, even though the dynamic is simple. We take the spatial translation into account by considering an alternative partition function which counts trajectories that end in cycles with a given spatial translation. Consider rule 2 shown in Fig. \ref{fig:shifting} (left). The rule does not have a cycle, but presents a simple repeating pattern that is translated left at each time step. In the case of a left-translated pattern, we consider the \textit{left-translated} neighborhoods shown in red in the figure. 
The partition function for this case reads
\begin{equation}\label{eq:partition function factorized translated left}
\begin{aligned}
Z_{\text{left}}=\sum_{\underline{\bm s}\in \underline{S}} &
\prod_{i=1}^N e^{ \mu s_1^i}
\mathds{1}\left[f(s_{p+c}^{i-1}, s_{p+c}^{i}, s_{p+c}^{i+1})=\bm s_{p+1}^{i-c}\right]
\\
&\times
\prod_{t=1}^{p+c-1}\mathds{1}\left[f(s_t^{i-1}, s_t^{i}, s_t^{i+1})=\bm s_{t+1}^i\right].
\end{aligned}
\end{equation}
The only difference with eq.\eqref{eq:partition function factorized} is the $i-c$ in the first indicator function, enforcing that the periodicity is now between site $i$ and $i-c$ and not site $i$ and itself. The transfer matrix associated with this new normalization is
\begin{equation}\label{eq:transfer_matrix_translated}
\begin{aligned}
\mathcal{T}^{\text{left}}_{(\underline{\lowers}^{\alpha}, \underline{\lowers}^{\beta}), (\underline{\sigma}^\alpha, \underline{\sigma}^{\beta})}
    = 
e^{ \mu s_1^i}
\mathds{1}\left[f(s_{p+c}^{\alpha}, s_{p+c}^{\beta}, \sigma_{p+c}^{\beta})= s_{p+1}^{\beta}\right]
\\
\times
\mathds{1}\left[\underline{s}^\beta=\underline{\sigma}^\alpha\right] \times
\prod_{t=1}^{p+c-1}\mathds{1}\left[f(s_t^{\alpha}, s_t^{\beta}, \sigma_t^{\beta})=\sigma_{t+1}^\beta\right],
\end{aligned}
\end{equation}
where the trajectories are \textit{left-translated}, i.e.
$\underline{s}^\alpha=\left(s^i_1, s^{i-1}_2, \hdots, s^{i-T+1}_T\right)$.
Note that in the product indicator functions of \eqref{eq:transfer_matrix_translated}, the translation is compensated by using $\sigma_{t+1}^\beta$, assuring the proper time evolution. Similarly, we can consider \textit{right-translated} neighborhoods by replacing $\sigma_{t+1}^\beta$ by $s^\alpha_{t+1}$ in the last indicator function. We also consider neighborhoods that do not shift for odd time-steps and shift for even time-steps (see Fig. \ref{fig:shifting} right).

\begin{figure}[t]
    \centering
    \begin{minipage}{0.2\textwidth}
        \centering
        \includegraphics[width=0.9\textwidth]{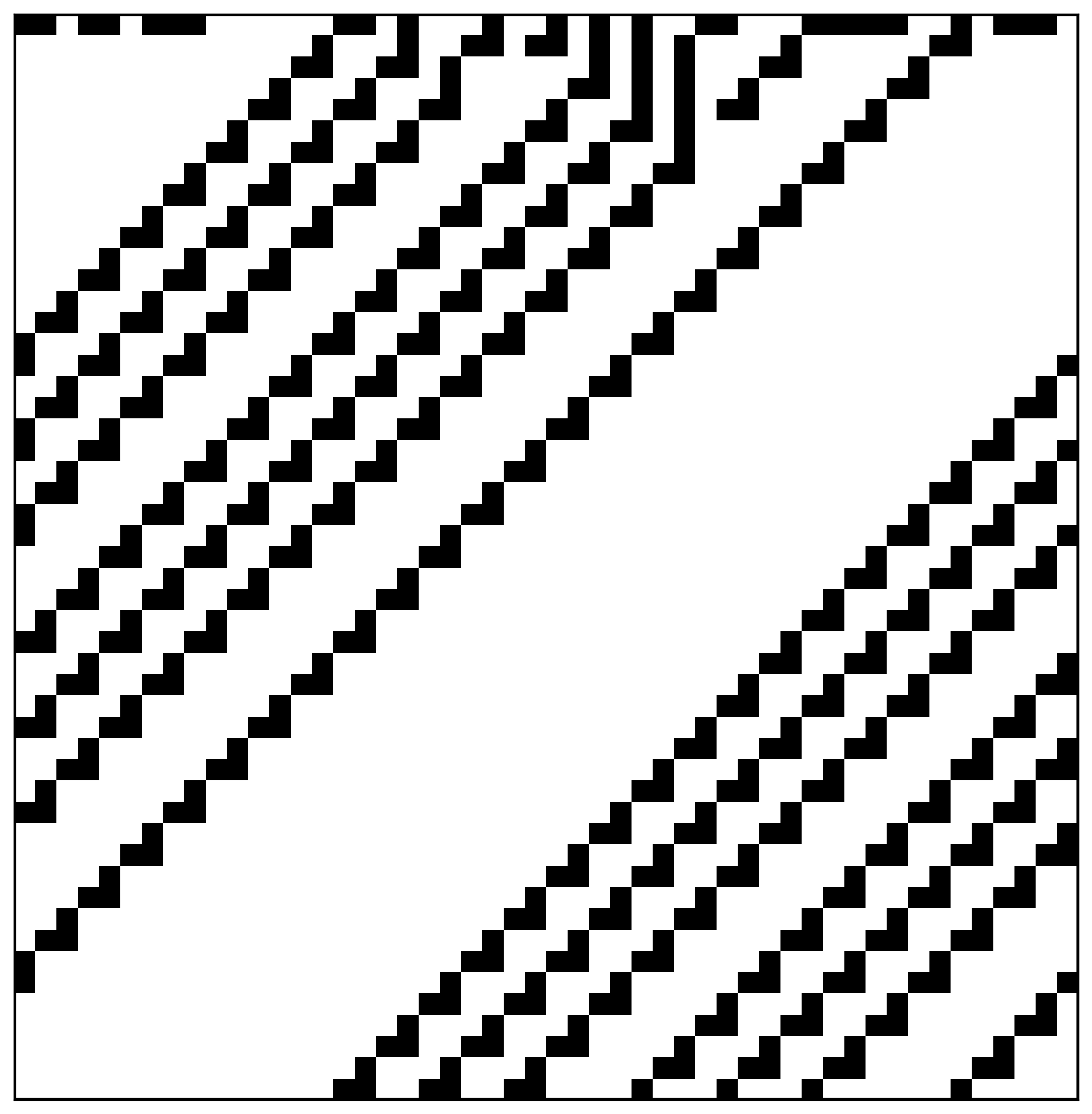}
    \end{minipage}
    \begin{minipage}{0.2\textwidth}
        \centering
        \includegraphics[width=0.9\textwidth]{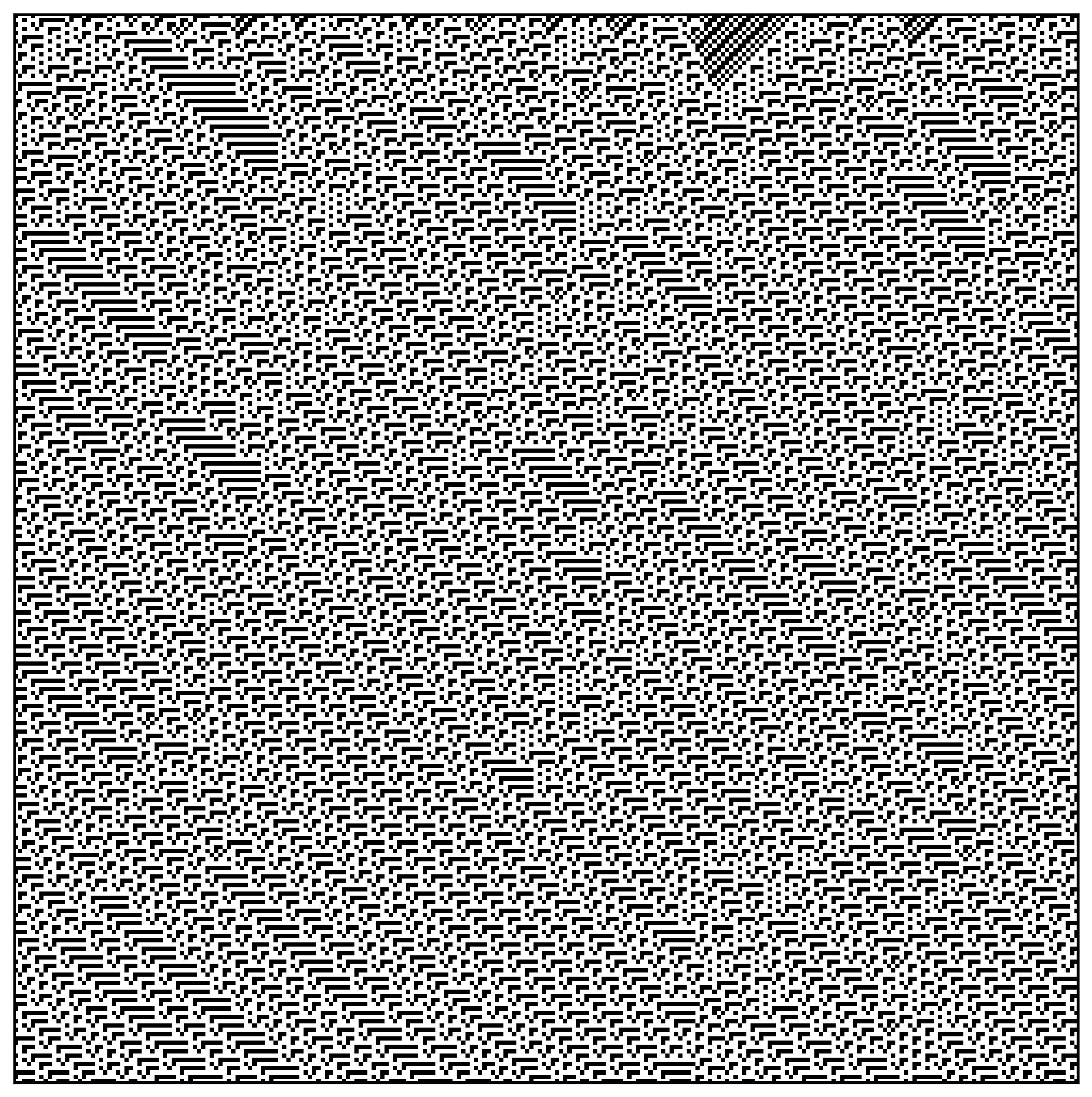}
    \end{minipage}

    \vspace{0.1em} 

    \begin{minipage}{0.2\textwidth}
        \centering
        \hspace{-0.17\textwidth} 
    \includegraphics[width=1\textwidth]{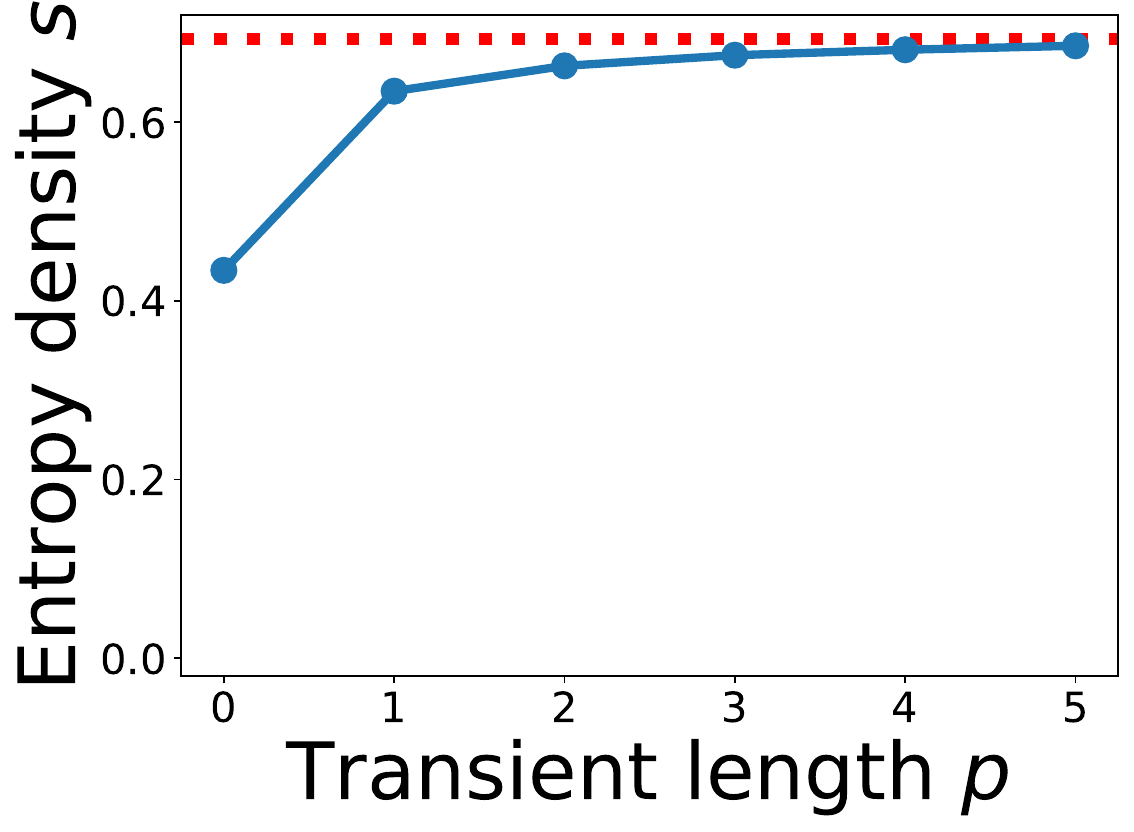}
    \end{minipage}
    \begin{minipage}{0.2\textwidth}
        \centering
        \hspace{-0.17\textwidth} 
    \includegraphics[width=1\textwidth]{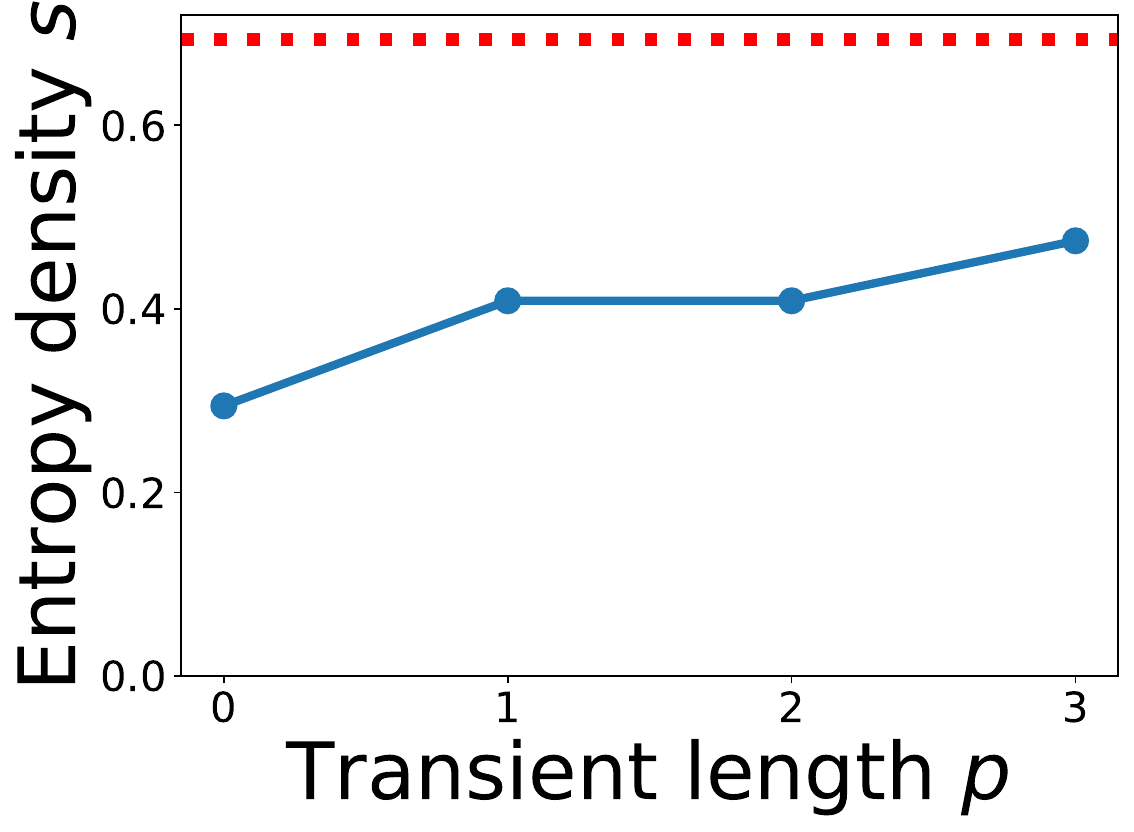}
    \end{minipage}

    \caption{Top: Space-time diagrams for rule 6 and rule 41. Bottom: Corresponding entropy profiles for $c=2$ and left-translated neighborhood (rule 6), $c=4$ and right-translated neighborhood (rule 41). We chose the value of $c$ and the neighborhood that presented the largest entropy. Note that the entropy of rule $6$ quickly reaches $\log(2)$, whereas the entropy of rule 41 increases slowly and is far from its maximum. These distinct entropy profiles align qualitatively with their space-time diagrams: rule 6 exhibits simpler periodic behavior versus the more elaborate structures of rule 41.}
    \label{fig:rule_6_41_comparison}
\end{figure}

With the appropriate choice of translation and $c$, the entropy increases rapidly in most cases (see Fig. \ref{fig:rule_6_41_comparison} left). However, some rules do not increase their entropy rapidly and are far from reaching $\log(2)$ within the number of time steps we tried. Observing the space-time diagrams of these rules, we notice some examples yield long-lived structures in their dynamics (see Fig. \ref{fig:rule_6_41_comparison} right), reminiscent of Class 4. Note that in \cite{castillo-ramirez_study_2025}, rule 41 is indeed classified as Class 4. Thus, the entropy growing but not reaching $\log(2)$ quickly as $p$ increases seems to signal the presence of these structures. We recall that the classification was established qualitatively on the visualization of the trajectories, so that some rules seem to be in between classes. This is reflected in the computed entropy values.

\vspace{-15pt} 

\begin{figure}
    \centering

     \centering
    \begin{minipage}{0.23\textwidth}
        \centering
        \hspace{-0.1\textwidth} 
      \includegraphics[width=0.9\textwidth]{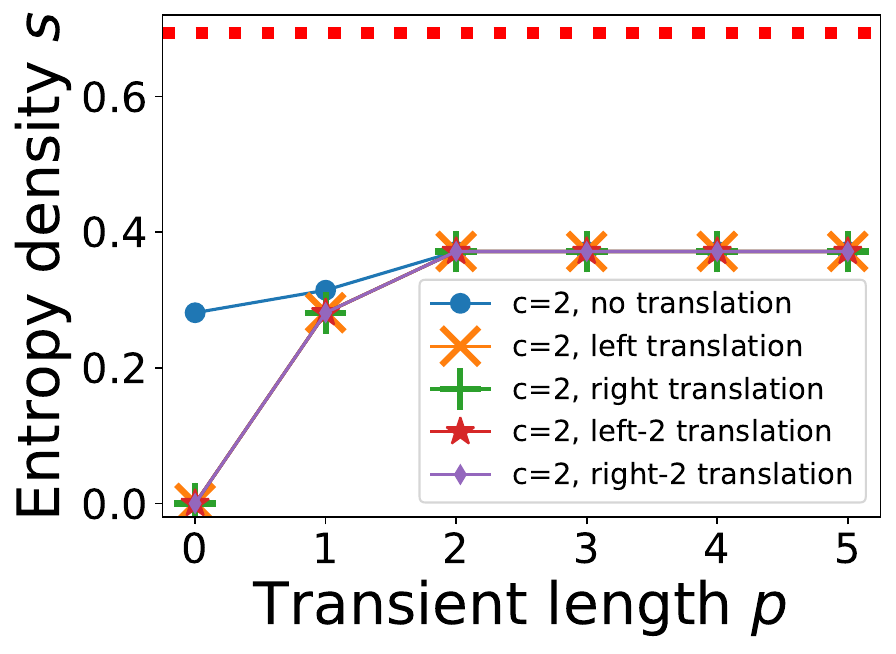}
    \end{minipage}
    \begin{minipage}{0.23\textwidth}
        \centering
        \hspace{-0.1\textwidth} 
\includegraphics[width=0.9\textwidth]{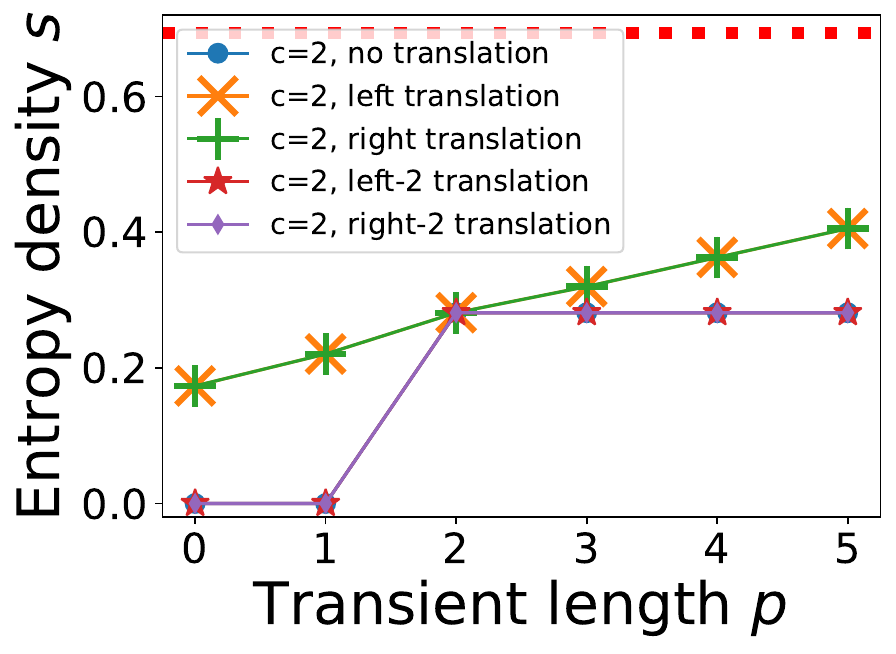}
    \end{minipage} 
    
    \caption{Entropy as function of $p$ for $c=2$ and different neighborhoods, rule 146 (left) and 54 (right). We note that changing the neighborhood does not change the entropy for $p>1$ for rule 146 (Class 3), whereas it does for rule 54 (Class 4). `left/right-2 translation' indicates a translation only at even time steps.}
    \label{fig:rule_146}
\end{figure}

\paragraph{Class 3}
Class 3 rules are described to lead to chaotic, quasi-random patterns. Thus, we expect the entropy density to remain low in a few steps. As the observed patterns seem random, we also expect the entropy not to change significantly when the neighborhood is translated. This is exemplified in Fig. \ref{fig:rule_146} (left) for rule 146. As an additional note, rules in Class 3 that have an entropy greater than $0$ for some choices of $p$, $c$ and neighborhood are in between classes in the classification of \cite{Hudcova_classification_2022}.

\vspace{-15pt} 

\paragraph{Class 4}
Class 4 rules present complex localized structures, sometimes long-lived. The computed entropies do not allow for a decisive distinction between Classes 3 and 4. However, we note that all Class 4 rules for ECAs have a positive entropy density, and that the entropy of rule 54 for left/right translated neighborhoods grows as $p$ increases, contrary to the other neighborhoods that plateau (see Fig. \ref{fig:rule_146} right). Rule 106 also presents different behavior depending on the neighborhood. The asymmetry observed between the neighborhoods (much less prevalent for all Class 3 rules) might be a sign of complexity, but the same phenomena was not found for rule 110.

\vspace{-10pt} 

\begin{figure}
    \centering
    \includegraphics[width=0.4\textwidth]{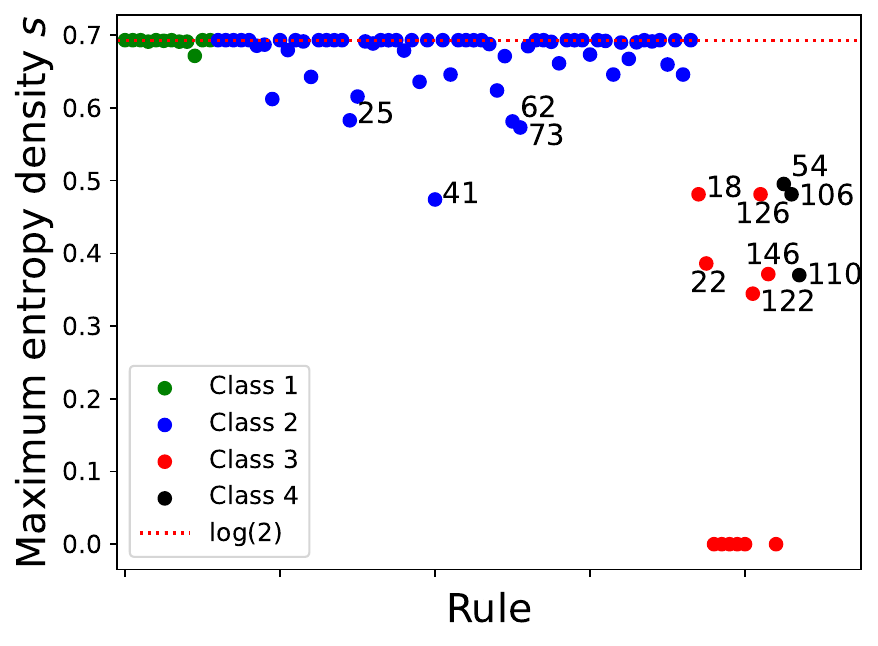}
    \caption{Maximum entropy density found for each rule among the tested $p$, $c$ and neighborhoods. The rules which are in between Classes 2 and 3 are drawn as Class 3. We indicate the Wolfram number for rules with intermediate entropies. We note the stark separation between Classes 1,2 and Classes 3,4. 
    }
    \label{fig:max_entropy_all_rules}
\end{figure}

\paragraph{Summary}
We computed the entropy density for the $88$ non-equivalent ECA rules. Fig. \ref{fig:max_entropy_all_rules} presents the maximum entropy density obtained among the combinations of $p=0,\hdots, 7$, $c=1,\hdots,5$ such that $p+c\leq 8$ for non-translated neighborhoods and $p+c\leq 7$ for translated neighborhoods. We notice the net separation between Classes 1, 2 and Classes 3, 4. The maximum obtained entropy is written in Table \ref{tab:rule-summary} for each rule, and the parameters used to obtain them. Note that the algorithm did not converge for certain parameter combinations after $10 \cdot 4^{p+c}$ iterations for rules 26, 30, 45, 60, 90, 105, 106, 150, 152, and 154. We do not expect these cases to yield higher entropies than those reported in the table, as a slowing down was noticed only for parameter values associated with negative or near-zero entropies. The baseline case $p=0, c=1$ that quantifies the number of fixed points (stationary configurations) is also reported in the table. This specific entropy is of independent interest as it characterizes the complexity of the simplest attractors, indicating whether a rule supports no stationary configurations ($s<0$), sub-exponentially many stationary configurations ($s=0$) or exponentially many stationary configurations ($s>0$). This complements \cite{koller_counting_2024}, where the entropy of stationary configurations of outer-totalistic graph cellular automata is computed. The complete table of results can be found in the github repository.

\begin{table*}[h!]
\centering
\small
\renewcommand{\arraystretch}{1.2}

\begin{minipage}[t]{0.48\linewidth}
\centering
\begin{tabularx}{\linewidth}{c c c c X}
\toprule
\textbf{Rule} & \textbf{Class} & $p=0$, $c=1$ & \textbf{Max} $s$ & $p$, $c$, \textbf{Neighborhood} \\
\midrule
0 & 1 & 0.000 & 0.693 & p=2, c=5, right-tr. \\
1 & 2 & $-\infty$ & 0.693 & p=5, c=2, no tr. \\
2 & 2 & 0.000 & 0.693 & p=1, c=1, left-tr. \\
3 & 2 & $-\infty$ & 0.693 & p=3, c=4, right-2 tr. \\
4 & 2 & 0.481 & 0.693 & p=2, c=5, no tr. \\
5 & 2 & 0.281 & 0.693 & p=1, c=2, no tr. \\
6 & 2 & 0.000 & 0.686 & p=5, c=2, left-tr. \\
7 & 2 & 0.000 & 0.687 & p=5, c=2, right-2 tr. \\
8 & 1 & 0.000 & 0.693 & p=6, c=1, right-2 tr. \\
9 & 2 & $-\infty$ & 0.612 & p=5, c=2, right-tr. \\
10 & 2 & 0.000 & 0.693 & p=1, c=1, left-2 tr. \\
11 & 2 & $-\infty$ & 0.680 & p=5, c=2, right-tr. \\
12 & 2 & 0.481 & 0.693 & p=6, c=2, no tr. \\
13 & 2 & 0.281 & 0.691 & p=7, c=1, no tr. \\
14 & 2 & 0.000 & 0.643 & p=6, c=1, left-tr. \\
15 & 2 & 0.000 & 0.693 & p=3, c=4, right-tr. \\
18 & 2/3 & 0.000 & 0.481 & p=4, c=3, right-tr. \\
19 & 2 & $-\infty$ & 0.693 & p=6, c=2, no tr. \\
22 & 2/3 & 0.000 & 0.386 & p=4, c=4 no tr. \\
23 & 2 & 0.000 & 0.693 & p=6, c=2, no tr. \\
24 & 2 & 0.000 & 0.693 & p=5, c=2, right-tr. \\
25 & 2 & $-\infty$ & 0.583 & p=5, c=2, right-2 tr. \\
26 & 2 & 0.000 & 0.616 & p=3, c=4, left-tr. \\
27 & 2 & $-\infty$ & 0.691 & p=5, c=2, right-2 tr. \\
28 & 2 & 0.000 & 0.689 & p=6, c=2, no tr. \\
29 & 2 & 0.000 & 0.693 & p=4, c=4, no tr. \\
30 & 3 & 0.000 & 0.000 & p=0, c=2, right-tr. \\
32 & 1 & 0.000 & 0.693 & p=7, c=1, no tr. \\
33 & 2 & $-\infty$ & 0.693 & p=6, c=2, no tr. \\
34 & 2 & 0.000 & 0.693 & p=2, c=3, left-tr. \\
35 & 2 & $-\infty$ & 0.679 & p=5, c=2, right-2 tr. \\
36 & 2 & 0.382 & 0.693 & p=6, c=1, no tr. \\
37 & 2 & 0.000 & 0.636 & p=6, c=2, no tr. \\
38 & 2 & 0.000 & 0.693 & p=5, c=2, left-tr. \\
40 & 1 & 0.000 & 0.691 & p=7, c=1, no tr. \\
41 & 2 & $-\infty$ & 0.474 & p=3, c=4, right-tr. \\
42 & 2 & 0.000 & 0.693 & p=1, c=1, left-tr. \\
43 & 2 & $-\infty$ & 0.646 & p=6, c=1, left-tr. \\
44 & 2 & 0.382 & 0.693 & p=7, c=1, no tr. \\
45 & 3 & 0.000 & 0.000 & p=3, c=3, right-tr. \\
46 & 2 & 0.000 & 0.693 & p=5, c=2, left-tr. \\
50 & 2 & 0.000 & 0.693 & p=6, c=2, no tr. \\
51 & 2 & $-\infty$ & 0.693 & p=5, c=2, no tr. \\
54 & 2/4 & 0.000 & 0.495 & p=4, c=4, no tr. \\
\bottomrule
\end{tabularx}
\end{minipage}
\hfill
\begin{minipage}[t]{0.48\linewidth}
\centering
\begin{tabularx}{\linewidth}{c c c c X}
\toprule
\textbf{Rule} & \textbf{Class} & $p=0$, $c=1$ & \textbf{Max} $s$ & $p$, $c$, \textbf{Neighborhood} \\
\midrule
56 & 2 & 0.000 & 0.688 & p=6, c=1, right-tr. \\
57 & 2 & $-\infty$ & 0.624 & p=6, c=1, left-tr. \\
58 & 2 & 0.000 & 0.671 & p=6, c=1, left-tr. \\
60 & 2/3 & 0.000 & 0.000 & p=5, c=2, no tr. \\
62 & 2 & 0.000 & 0.581 & p=5, c=3, no tr. \\
72 & 1 & 0.382 & 0.693 & p=6, c=1, right-2 tr. \\
73 & 2/3/4 & 0.199 & 0.573 & p=6, c=2, no tr. \\
74 & 2 & 0.000 & 0.685 & p=5, c=2, left-tr. \\
76 & 2 & 0.609 & 0.693 & p=5, c=1, no tr. \\
77 & 2 & 0.481 & 0.693 & p=7, c=1, no tr. \\
78 & 2 & 0.281 & 0.691 & p=7, c=1, no tr. \\
90 & 2/3 & 0.000 & 0.000 & p=3, c=1, left-2 tr. \\
94 & 2 & 0.281 & 0.661 & p=6, c=2, no tr. \\
104 & 1 & 0.322 & 0.692 & p=7, c=1, no tr. \\
105 & 2/3 & 0.000 & 0.000 & p=1, c=2, right-tr. \\
106 & 3/4 & 0.000 & 0.481 & p=6, c=1, left-tr. \\
108 & 2 & 0.481 & 0.693 & p=4, c=4, no tr. \\
110 & 4 & 0.000 & 0.370 & p=3, c=3, right-2 tr. \\
122 & 2/3 & 0.000 & 0.345 & p=6, c=1, right-tr. \\
126 & 2/3 & 0.000 & 0.481 & p=7, c=1, no tr. \\
128 & 1 & 0.000 & 0.693 & p=7, c=1, no tr. \\
130 & 2 & 0.000 & 0.693 & p=6, c=1, left-tr. \\
132 & 2 & 0.481 & 0.693 & p=7, c=1, no tr. \\
134 & 2 & 0.000 & 0.673 & p=5, c=2, left-tr. \\
136 & 1 & 0.000 & 0.691 & p=7, c=1, no tr. \\
138 & 2 & 0.000 & 0.693 & p=2, c=1, left-tr. \\
140 & 2 & 0.481 & 0.692 & p=7, c=1, no tr. \\
142 & 2 & 0.000 & 0.646 & p=6, c=1, left-tr. \\
146 & 2/3 & 0.000 & 0.371 & p=7, c=1, no tr. \\
150 & 2/3 & 0.000 & 0.000 & p=3, c=1, right-2 tr. \\
152 & 2 & 0.000 & 0.690 & p=6, c=1, right-tr. \\
154 & 2/3 & 0.000 & 0.667 & p=0, c=4, left-tr. \\
156 & 2 & 0.000 & 0.690 & p=6, c=2, no tr. \\
160 & 1 & 0.000 & 0.691 & p=7, c=1, no tr. \\
162 & 2 & 0.000 & 0.693 & p=6, c=1, left-tr. \\
164 & 2 & 0.382 & 0.691 & p=7, c=1, no tr. \\
168 & 1 & 0.000 & 0.672 & p=7, c=1, no tr. \\
170 & 2 & 0.000 & 0.693 & p=6, c=1, left-tr. \\
172 & 2 & 0.382 & 0.660 & p=7, c=1, no tr. \\
178 & 2 & 0.000 & 0.693 & p=6, c=2, no tr. \\
184 & 2 & 0.000 & 0.646 & p=6, c=1, right-tr. \\
200 & 1 & 0.562 & 0.693 & p=1, c=5, no tr. \\
204 & 2 & 0.693 & 0.693 & p=6, c=1, right-2 tr. \\
232 & 1 & 0.481 & 0.693 & p=7, c=1, no tr. \\
\bottomrule
\end{tabularx}
\end{minipage}

\caption{Summary of the obtained entropy densities. The second column indicates the Wolfram class. The third column indicates the entropy of stationary configurations. The fourth column indicates the maximum entropy obtained among all the tested values of $p$, $c$ and neighborhood. The last column indicates for which of these values the maximum entropy is obtained. It can be that multiple set of parameters lead to the same maximal entropy, but only one set is indicated in this table. `tr.' indicates `translation', and `left/right-2 tr.' indicates a translation only at even time steps. Numerical estimates of large negative entropy ($s<-5$) are indicated as $-\infty$, to signify that there are no initial trajectory respecting the constraints.
}
\label{tab:rule-summary}
\end{table*}


\section{Conclusion}

We introduced a method based on the Transfer Matrix formalism to compute the entropy density of trajectories in Elementary Cellular Automata. This approach provides an exact quantitative measure, in the thermodynamic limit, of the number of initial conditions leading to specific dynamical outcomes (a $p$-step transient followed by a $c$-step cycle). Our results establish a clear connection between this trajectory entropy and the qualitative Wolfram classification. Class 1 and 2 rules exhibit high entropy that rapidly approaches the theoretical maximum as the transient length $p$ increases, reflecting their simple dynamics and weak correlation on initial states. In contrast, Class 3 and 4 rules show significantly lower entropy values (often zero, or positive but quickly saturating with $p$), indicative of chaotic or complex evolution. While limited computationally to short trajectories, this method offers a valuable tool for probing the statistical properties of CA dynamics and provides a quantitative basis for understanding the emergence of different dynamical regimes from simple local rules.

The formalism presented here, while focused on entropy, enables the study of other observables (as demonstrated with the initial density), suggesting avenues for future exploration. Dedicated studies of additional observables could yield insights into phenomena such as dynamical phase transitions \citep{behrens_dynamical_2023}. The framework also lends itself to optimization approaches; for instance, algorithms could potentially be designed to identify initial configurations yielding trajectories with low entropy, thereby isolating states with specific structural or dynamical properties. Extending these computations to larger timescales remains a significant area for future work, which will probably necessitate the development of appropriate approximation techniques to address our method's inherent computational scaling. Furthermore, while 1D, sparse and dense systems can be analyzed using well-established methods, the analysis of 2D and 3D lattice models remains challenging, even in the static case \citep{Baxter_Exactly_1984, viswanathan2022does, duminil2022100}. These complexities highlight the need for further exploration and development of more robust techniques to handle such systems in future studies.

\section{Acknowledgements}
This project originated as part of the Artificial Life course (MATH-642) offered in Spring 2024 at École Polytechnique Fédérale de Lausanne (EPFL). The authors are grateful to Vassilis Papadopoulos for valuable discussions and insightful feedback. The project was funded by NCCR SwissMAP.

\footnotesize
\bibliographystyle{apalike}
\bibliography{example} 

\begin{thebibliography}{}

\bibitem[Alfaro and Sanjuán, 2024]{alfaro_classification_2024}
Alfaro, G. and Sanjuán, M. A.~F. (2024).
\newblock Classification of cellular automata based on the hamming distance.
\newblock {\em Chaos: An Interdisciplinary Journal of Nonlinear Science}, 34(8):083129.

\bibitem[Baxter, 1984]{Baxter_Exactly_1984}
Baxter, R.~J. (1984).
\newblock {\em Exactly Solved Models in Statistical Mechanics}, pages 5--63.

\bibitem[Behrens et~al., 2023]{behrens_backtracking_2023}
Behrens, F., Hudcov\'a, B., and Zdeborov\'a, L. (2023).
\newblock Backtracking dynamical cavity method.
\newblock {\em Phys. Rev. X}, 13:031021.

\bibitem[Behrens et~al., 2024]{behrens_dynamical_2023}
Behrens, F., Hudcov\'a, B., and Zdeborov\'a, L. (2024).
\newblock Dynamical phase transitions in graph cellular automata.
\newblock {\em Phys. Rev. E}, 109:044312.

\bibitem[Bhattacharjee et~al., 2020]{bhattacharjee_survey_2020}
Bhattacharjee, K., Naskar, N., Roy, S., and Das, S. (2020).
\newblock A survey of cellular automata: types, dynamics, non-uniformity and applications.
\newblock {\em Natural Computing}, 19(2):433--461.

\bibitem[Blanchard et~al., 1997]{blanchard_topological_1997}
Blanchard, F., Kůrka, P., and Maass, A. (1997).
\newblock Topological and measure-theoretic properties of one-dimensional cellular automata.
\newblock {\em Physica D: Nonlinear Phenomena}, 103(1):86--99.

\bibitem[Castillo-Ramirez and Maga{\~n}a-Chavez, 2025]{castillo-ramirez_study_2025}
Castillo-Ramirez, A. and Maga{\~n}a-Chavez, M.~G. (2025).
\newblock A study on the composition of elementary cellular automata.
\newblock In {\em Advances in Cellular Automata: Volume 1: Theory}, pages 347--373. Springer.

\bibitem[Chopard and Droz, 1998]{chopard_cellular_1998}
Chopard, B. and Droz, M. (1998).
\newblock {\em Cellular Automata Modeling of Physical Systems}.
\newblock Collection Alea-Saclay: Monographs and Texts in Statistical Physics. Cambridge University Press.

\bibitem[Cook, 2004]{cook_universality}
Cook, M. (2004).
\newblock Universality in elementary cellular automata.
\newblock {\em Complex Systems}, 15.

\bibitem[Coolen and Takeda, 2012]{coolen_transfer_2012}
Coolen, A.~C. and Takeda, K. (2012).
\newblock Transfer operator analysis of the parallel dynamics of disordered ising chains.
\newblock {\em Philosophical Magazine}, 92(1):64--77.

\bibitem[Culik and Yu, 1988]{culik1988undecidability}
Culik, K. and Yu, S. (1988).
\newblock Undecidability of ca classification schemes.
\newblock {\em Complex Systems}, 2(2):177--190.

\bibitem[Delacourt, 2021]{delacourt_2021}
Delacourt, M. (2021).
\newblock {Rice’s Theorem for Generic Limit Sets of Cellular Automata}.
\newblock In {\em 27th IFIP WG 1.5 International Workshop on Cellular Automata and Discrete Complex Systems (AUTOMATA 2021)}, volume~90 of {\em Open Access Series in Informatics (OASIcs)}, pages 6:1--6:12.

\bibitem[Duminil-Copin, 2022]{duminil2022100}
Duminil-Copin, H. (2022).
\newblock 100 years of the (critical) ising model on the hypercubic lattice.
\newblock In {\em Proceedings of the International Congress of Mathematicians}, volume~1, pages 164--210.

\bibitem[Ermentrout and Edelstein-Keshet, 1993]{ermentrout_cellular_1993}
Ermentrout, G.~B. and Edelstein-Keshet, L. (1993).
\newblock Cellular automata approaches to biological modeling.
\newblock {\em J Theor Biol}, 160(1):97--133.

\bibitem[Horn and Johnson, 2012]{horn2012matrix}
Horn, R.~A. and Johnson, C.~R. (2012).
\newblock {\em Matrix analysis}.
\newblock Cambridge university press.

\bibitem[Hudcová and Mikolov, 2022]{Hudcova_classification_2022}
Hudcová, B. and Mikolov, T. (2022).
\newblock Classification of discrete dynamical systems based on transients.
\newblock {\em Artificial Life}, 27(3–4):220--245.

\bibitem[Kari, 1994a]{kari1994reversibility}
Kari, J. (1994a).
\newblock Reversibility and surjectivity problems of cellular automata.
\newblock {\em Journal of Computer and System Sciences}, 48(1):149--182.

\bibitem[Kari, 1994b]{kari_rices_1994}
Kari, J. (1994b).
\newblock Rice's theorem for the limit sets of cellular automata.
\newblock {\em Theoretical Computer Science}, 127(2):229--254.

\bibitem[Koller et~al., 2024]{koller_counting_2024}
Koller, C., Behrens, F., and Zdeborov{\'a}, L. (2024).
\newblock Counting and hardness-of-finding fixed points in cellular automata on random graphs.
\newblock {\em Journal of Physics A: Mathematical and Theoretical}, 57(46):465001.

\bibitem[Kramers and Wannier, 1941a]{kramers_statistics_1941_1}
Kramers, H.~A. and Wannier, G.~H. (1941a).
\newblock Statistics of the two-dimensional ferromagnet. part i.
\newblock 60(3):252--262.
\newblock Publisher: American Physical Society.

\bibitem[Kramers and Wannier, 1941b]{kramers_statistics_1941_2}
Kramers, H.~A. and Wannier, G.~H. (1941b).
\newblock Statistics of the two-dimensional ferromagnet. part ii.
\newblock 60(3):263--276.
\newblock Publisher: American Physical Society.

\bibitem[Langton, 1990]{langton1990computation}
Langton, C.~G. (1990).
\newblock Computation at the edge of chaos: Phase transitions and emergent computation.
\newblock {\em Physica D: nonlinear phenomena}, 42(1-3):12--37.

\bibitem[Lehoucq et~al., 1998]{ARPACK}
Lehoucq, R.~B., Sorensen, D.~C., and Yang, C. (1998).
\newblock 1. introduction to {ARPACK}.
\newblock In {\em {ARPACK} Users' Guide}, Software, Environments, and Tools, pages 1--7. Society for Industrial and Applied Mathematics.

\bibitem[Lemoy et~al., 2014]{lemoy_transfer_2014}
Lemoy, R., Mozeika, A., and Seki, S. (2014).
\newblock Transfer matrix analysis of one-dimensional majority cellular automata with thermal noise.
\newblock {\em J. Phys. A: Math. Theor.}, 47(10):105001.
\newblock Publisher: {IOP} Publishing.

\bibitem[Li and Packard, 1990]{li1990structure}
Li, W. and Packard, N. (1990).
\newblock The structure of the elementary cellular automata rule space.
\newblock {\em Complex systems}, 4(3):281--297.

\bibitem[L{\'o}pez-D{\'\i}az et~al., 2023]{lopez2023temporal}
L{\'o}pez-D{\'\i}az, A.~J., S{\'a}nchez-Puig, F., and Gershenson, C. (2023).
\newblock Temporal, structural, and functional heterogeneities extend criticality and antifragility in random boolean networks.
\newblock {\em Entropy}, 25(2):254.

\bibitem[Marr and Hütt, 2005]{marr_topology_2005}
Marr, C. and Hütt, M.-T. (2005).
\newblock Topology regulates pattern formation capacity of binary cellular automata on graphs.
\newblock {\em Physica A: Statistical Mechanics and its Applications}, 354:641--662.
\newblock Publisher: Elsevier.

\bibitem[Marr and Hütt, 2012]{marr_cellular_2012}
Marr, C. and Hütt, M.-T. (2012).
\newblock Cellular automata on graphs: Topological properties of {ER} graphs evolved towards low-entropy dynamics.
\newblock {\em Entropy}, 14(6):993--1010.
\newblock Number: 6 Publisher: Molecular Diversity Preservation International.

\bibitem[Martinez, 2013]{martinez_note}
Martinez, G.~J. (2013).
\newblock A note on elementary cellular automata classification.
\newblock {\em Journal of cellular automata}, 8.

\bibitem[Prokopenko et~al., 2009]{prokopenko2009information}
Prokopenko, M., Boschetti, F., and Ryan, A.~J. (2009).
\newblock An information-theoretic primer on complexity, self-organization, and emergence.
\newblock {\em Complexity}, 15(1):11--28.

\bibitem[Santamar{\'\i}a-Bonfil et~al., 2017]{santamaria2017package}
Santamar{\'\i}a-Bonfil, G., Gershenson, C., and Fern{\'a}ndez, N. (2017).
\newblock A package for measuring emergence, self-organization, and complexity based on shannon entropy.
\newblock {\em Frontiers in Robotics and AI}, 4:10.

\bibitem[Schaller and Svozil, 2025]{SCHALLER2025100298}
Schaller, M. and Svozil, K. (2025).
\newblock Orbits of one-dimensional cellular automata induced by symmetry transformations.
\newblock {\em Physics Open}, 24:100298.

\bibitem[Sorensen, 1992]{sorensen_implicit_1992}
Sorensen, D.~C. (1992).
\newblock Implicit application of polynomial filters in a k-step arnoldi method.
\newblock {\em {SIAM} J. Matrix Anal. Appl.}, 13(1):357--385.
\newblock Publisher: Society for Industrial and Applied Mathematics.

\bibitem[Sorensen, 1997]{sorensen_implicitly_1997}
Sorensen, D.~C. (1997).
\newblock Implicitly restarted arnoldi/lanczos methods for large scale eigenvalue calculations.
\newblock In Keyes, D.~E., Sameh, A., and Venkatakrishnan, V., editors, {\em Parallel Numerical Algorithms}, pages 119--165. Springer.

\bibitem[Virtanen et~al., 2020]{2020SciPy-NMeth}
Virtanen, P., Gommers, R., Oliphant, T.~E., Haberland, M., Reddy, T., Cournapeau, D., Burovski, E., Peterson, P., Weckesser, W., Bright, J., {van der Walt}, S.~J., Brett, M., Wilson, J., Millman, K.~J., Mayorov, N., Nelson, A. R.~J., Jones, E., Kern, R., Larson, E., Carey, C.~J., Polat, {\.I}., Feng, Y., Moore, E.~W., {VanderPlas}, J., Laxalde, D., Perktold, J., Cimrman, R., Henriksen, I., Quintero, E.~A., Harris, C.~R., Archibald, A.~M., Ribeiro, A.~H., Pedregosa, F., {van Mulbregt}, P., and {SciPy 1.0 Contributors} (2020).
\newblock {{SciPy} 1.0: Fundamental Algorithms for Scientific Computing in Python}.
\newblock {\em Nature Methods}, 17:261--272.

\bibitem[Vispoel et~al., 2022]{vispoel_progress_2022}
Vispoel, M., Daly, A.~J., and Baetens, J.~M. (2022).
\newblock Progress, gaps and obstacles in the classification of cellular automata.
\newblock {\em Physica D: Nonlinear Phenomena}, 432:133074.

\bibitem[Viswanathan et~al., 2022]{viswanathan2022does}
Viswanathan, G.~M., Portillo, M. A.~G., Raposo, E.~P., and da~Luz, M.~G. (2022).
\newblock What does it take to solve the 3d ising model? minimal necessary conditions for a valid solution.
\newblock {\em Entropy}, 24(11):1665.

\bibitem[Wolfram, 1983]{wolfram_statistical_1983}
Wolfram, S. (1983).
\newblock Statistical mechanics of cellular automata.
\newblock {\em Rev. Mod. Phys.}, 55(3):601--644.
\newblock Publisher: American Physical Society.

\bibitem[Wolfram, 1984]{wolfram_universality_1984}
Wolfram, S. (1984).
\newblock Universality and complexity in cellular automata.
\newblock {\em Physica D: Nonlinear Phenomena}, 10(1):1--35.

\bibitem[Wuensche and Lesser, 2001]{Wuensche_global_2001}
Wuensche, A. and Lesser, M. (2001).
\newblock The global dynamics of celullar automata: An atlas of basin of attraction fields of one-dimensional cellular automata.
\newblock {\em J. Artificial Societies and Social Simulation}, 4.

\bibitem[Yeomans, 1992]{yeomans_statistical_1992}
Yeomans, J.~M. (1992).
\newblock {\em Statistical Mechanics of Phase Transitions}.
\newblock Oxford University Press.

\bibitem[Zenil, 2010]{zenil_compression-based_2010}
Zenil, H. (2010).
\newblock Compression-based investigation of the dynamical properties of cellular automata and other systems.
\newblock {\em Complex Systems}, 19(1).

\end{thebibliography}

\end{document}